\documentclass[twocolumn,preprintnumbers,superscriptaddress,amsmath,amssymb,prb]{revtex4-1}

\usepackage{graphicx}
\usepackage{dcolumn}
\usepackage{subfigure}
\usepackage{bm}
\usepackage[utf8]{inputenc}
\usepackage{comment}
\usepackage[unicode=true, bookmarks=false, breaklinks=false, pdfborder={0 0 1},colorlinks=false]{hyperref}
\usepackage{breakurl}
\usepackage{url}
\usepackage{natbib}
\usepackage{multirow}
\usepackage{float}

\newcommand{\unit}[2][1]{#1~\mathrm{#2}}

\begin{document}

\title{Dzyaloshinskii-Moriya interactions, N\'eel skyrmions and V$_4$ magnetic clusters in multiferroic lacunar spinel GaV$_4$S$_8$}

\newcommand{\Uppsala}{Department of Physics and Astronomy, Uppsala University, Box 516, SE-75120 Uppsala, Sweden}
\newcommand{\KTH}{Department of Applied Physics, School of Engineering Sciences, KTH Royal Institute of Technology, 
AlbaNova University Center, SE-10691 Stockholm, Sweden}
\newcommand{\SeRC}{SeRC (Swedish e-Science Research Center), KTH Royal Institute of Technology, SE-10044 Stockholm, Sweden}

\author{Vladislav Borisov}
    \affiliation{\Uppsala}
    \email[Corresponding author:\ ]{vladislav.borisov@physics.uu.se}

\author{Nastaran Salehi}
    \affiliation{\Uppsala}

\author{Manuel Pereiro}
    \affiliation{\Uppsala}

\author{Anna Delin}
    \affiliation{\KTH}
    \affiliation{\SeRC}

\author{Olle Eriksson}
    \affiliation{\Uppsala}
    
\date{\today}

\begin{abstract}
Using \textit{ab initio} density functional theory with static mean-field correlations, we calculate the Heisenberg and Dzyaloshinskii-Moriya interactions (DMI) for an atomistic spin Hamiltonian for the lacunar spinel, GaV$_4$S$_8$. The parameters describing these interactions are used in atomistic spin dynamics and micromagnetic simulations. The magnetic properties of the lacunar spinel GaV$_4$S$_8$, a material well-known from experiment to host magnetic skyrmions of N\'eel character, are simulated with these \textit{ab initio} calculated parameters. The Dzyaloshinskii-Moriya contribution to the micromagnetic energy is a sum of two Lifshitz invariants, supporting the formation of N\'eel skyrmions and its symmetry agrees with what is usually expected for $C_{3\nu}$-symmetric systems. The are several conclusions one may draw from this work. One concerns the quantum nature of the magnetism, where we show that the precise magnetic state of the V$_4$ cluster is crucial for understanding quantitatively the magnetic phase diagram. In particular we demonstrate that a distributed-moment state of each V$_4$ cluster explains well a variety of properties of GaV$_4$S$_8$, such as the band gap, observed Curie temperature and especially the stability of N\'eel skyrmions in the experimentally relevant temperature and magnetic-field range. In addition, we find that electronic correlations change visibly the calculated value of the DMI.
\end{abstract}

\maketitle

\section{Introduction}

Magnetic skyrmions, which are topological spin textures, have been found in a few materials classes (B20 compounds [\onlinecite{Muehlbauer2009},\onlinecite{Yu2011}], Co-Mn-Zn alloys etc., see also a review in Ref.~[\onlinecite{Kanazawa2017}]) as well as in low-dimensional systems of transition metal multilayers (Pt/Co/Ta [\onlinecite{Wang2019}], Ir/Fe/Co/Pt [\onlinecite{Soumyanarayanan2017}], Pd/Fe/Ir(111) [\onlinecite{Romming2013}] etc.) where, for some systems, topological magnetism was observed even at room temperature and in the absence of applied magnetic field. Even more unique are bulk magnets where skyrmions coexist with ferroelectricity and the only known examples are Cu$_2$OSeO$_3$ and the lacunar spinels GaV$_4$S$_8$ [\onlinecite{Kezsmarki2015}], GaV$_4$Se$_8$ [\onlinecite{Bordacs2017},\onlinecite{Fujima2017}], GaMo$_4$S$_8$ [\onlinecite{Butykai2022}] and GaMo$_4$Se$_8$ [\onlinecite{Schueller2020}]. The spinel compounds are especially interesting, because the ferroelectricity is of rare orbital-driven origin and has a considerable magnitude and because the skyrmions are of N\'eel character. This is in contrast to all the other bulk systems, where only Bloch skyrmions are observed.
\begin{figure}
{\centering
\includegraphics[width=0.9\columnwidth]{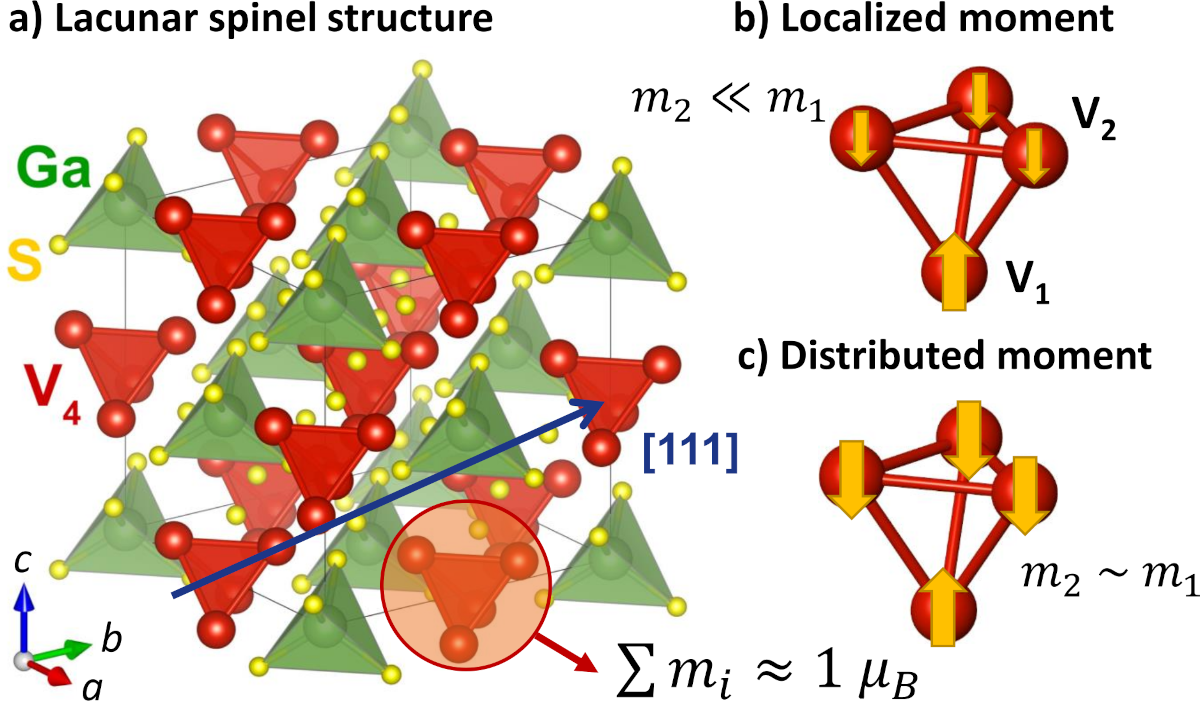}
}
\caption{a) Crystal structure of lacunar spinel GaV$_4$S$_8$ with V$_4$ clusters as the main magnetic units with effective spin $S=1/2$. The [111] axis (blue arrow) is shown, which corresponds to the $C_{3}$ rotation axis of the $C_{3\nu}$point group. b,c) The V$_4$ cluster of lacunar spinels in the low-temperature phase, where the distance between inequivalent V$_1$ and V$_2$ sites is $\unit[2.9]{\AA}$ and $d(\mathrm{V}_2-\mathrm{V}_2)=\unit[2.8]{\AA}$. The yellow arrows show the magnetic moments of individual V sites ($m_1$ for V$_1$ and $m_2$ for V$_2$) in b) the localized-moment (``L'' state, $m_2 \ll m_1$) and c) the distributed-moment (``D'' state $m_2 \sim m_1$) configurations. For details, see the main text.}
\vspace{-20pt}
\label{f:spinel_structure}
\end{figure}

Such a unique behavior of lacunar spinels has been attributed to the $C_{3\nu}$ point group of the crystal structure (Fig.~\ref{f:spinel_structure}). It was argued [\onlinecite{Bogdanov2002}] that this symmetry implies a specific form of the Lifshitz invariants describing the Dzyaloshinskii-Moriya interaction (DMI), which leads to the stability of N\'eel skyrmions and contrasts with the isotropic DMI in B20 compounds where only Bloch skyrmions emerge. At the same time, theoretical studies of the DMI in lacunar spinels are sparse. In many papers, some values of magnetic interactions between $S=1/2$ V$_4$-clusters are assumed and spin models with only nearest neighbors are used then to model the magnetic textures at varying external magnetic field strengths[\onlinecite{Kezsmarki2015},\onlinecite{Fujima2017}]. On the other hand, there are a few works where the Heisenberg and DM interactions between the V$_4$-clusters are actually calculated, using perturbation theory or total-energy fitting methods [\onlinecite{Zhang2017,Nikolaev2019,Nikolaev2020,Schueller2020}], and the results clarify the symmetry of DM vectors and the relative energy scales of different interactions in the system. Neutron experiments [\onlinecite{Dally2020}] and theoretical studies [\onlinecite{Schueller2019}] propose that the magnetization is uniformly distributed over all V sites in the V$_4$ cluster. Nevertheless, it is not clear yet how the DMI in lacunar spinels is affected by electronic correlations and details of the magnetic state of the four-site transition metal clusters, which play the role of effective spins and form a face-centered network (Fig.~\ref{f:spinel_structure}). The work presented here is aimed at closing this gap by a systematic analysis of electronic and magnetic properties of the first skyrmionic lacunar spinel GaV$_4$S$_8$.

\section{Theoretical methods}

To understand the magnetic phenomena in GaV$_4$S$_8$ we follow a multiscale approach where we start with a description on the level of individual electrons, proceed with atomistic magnetic interactions between effective V$_4$ cluster moments and, finally, model the magnetic textures at finite temperature in external field on length scales in the range $\unit[10-10^3]{nm}$. Details about these three main steps are given further below.

\vspace{5pt}
\textbf{Step 1 (electronic properties).} The electronic structure and magnetic properties of lacunar spinels are studied in this work using density functional theory (DFT)~[\onlinecite{Hohenberg1964}] in the full-potential Linear Muffin-Tin Orbital implementation, available in the RSPt code [\onlinecite{Wills1987},\onlinecite{Wills2010}]. Electronic correlations are modeled here on the static mean-field level by means of the DFT+$U$ approach with varied $U$ and fixed Hund's coupling $J_\mathrm{H}=\unit[0.9]{eV}$ on top of the spin-polarized local-density or generalized-gradient approximations of the DFT. Summation in the Brillouin zone is performed on the shifted $(16\times 16\times 16)$ $k$-mesh and the Fermi smearing with a temperature of $\unit[1]{mRy}$ is used for electronic occupations. All calculations are performed for the known experimental structure reported in the literature.

Two different V$_4$ cluster configurations are considered here, where the magnetic moment is either localized mostly on one V site or distributed over the whole cluster (four V sites, see Fig. \ref{f:spinel_structure}c). Note that the total moment per cluster is around $\unit[1]{\mu_\mathrm{B}}$ in both cases. It also has to be noted that, because of the elongation of V$_4$ tetrahedra along the [111]-direction, one of the V sites (V$_1$) is not symmetry-related to the other three sites (V$_2$) which are, however, equivalent to each other. As will become clear in the following, the cluster configuration can change the calculated properties substantially.

As our on-going calculations suggest and in accordance with literature [\onlinecite{Schueller2020}], the other lacunar spinel GaMo$_4$Se$_8$ with 4\textit{d} states shows more uniformly magnetized Mo$_4$ clusters with parallel-aligned spin moments, in contrast to the 3\textit{d} V-based spinels. This may indicate a fundamental difference between the 3\textit{d} and 4\textit{d} lacunar spinels, which will be discussed in a future work.

\vspace{5pt}
\textbf{Step 2 (magnetic exchange).} Magnetic interactions are calculated using the well-established Lichtenstein-Katsnelson-Antropov-Gubanov (LKAG) approach [\onlinecite{LKAG1987}] (for a recent review see [\onlinecite{Jijreview2023}]), where the idea is to relate the interaction between two spins to the energy change due to a small perturbation of the magnetic state. We use the muffin-tin projection to calculate site-specific electronic parameters [\onlinecite{Kvashnin2015}] and restrict ourselves to bilinear contributions to the magnetic energy for each pair of spins $\vec{S}_i$, which is written as follows:
\begin{equation}
    H = -J_{ij} (\vec{S}_i \cdot \vec{S}_j) -\vec{D}_{ij} \cdot (\vec{S}_i \times \vec{S}_j) - \vec{S}_i\, \hat\Gamma_{ij}\, \vec{S}_j,
    \label{e:Heisenberg_model}
\end{equation}
where one can distinguish between the 
isotropic Heisenberg ($J_{ij}$), Dzyaloshinskii-Moriya ($\vec{D}_{ij}$) and symmetric anisotropic ($\hat\Gamma_{ij}$) exchange interactions. 
To calculate the DM interaction, we perform three independent calculations where the spin axis is oriented along the $x$-, $y$- and $z$-directions to obtain the $D_x$, $D_y$ and $D_z$ components. It is also worth mentioning that the total magnetic moment of GaV$_4$S$_8$ varies only slightly upon such a global rotation of the magnetization. We also find that the symmetric anisotropic exchange $\hat\Gamma_{ij}$ is one or two orders of magnitude smaller than the DM interaction for different bonds, so we do not include this type of exchange in further simulations for GaV$_4$S$_8$. Application of the whole approach described above to several transition metal systems in our previous works [\onlinecite{Borisov2021,Ntallis2021,Borisov2022}] has demonstrated the reliability of the calculated values of magnetic interactions, which justifies the use of this approach in the present work.

The critical temperature $T_c$ for the magnetic ordering is estimated from Monte Carlo simulations based on the calculated exchange interactions $J_{ij}$ and $\vec{D}_{ij}$. Simulations are done using the \textsc{UppASD} code~[\onlinecite{uppasd},\onlinecite{Eriksson2017}] for bulk supercells containing $(N\times N\times N)$ unit cells with periodic boundary conditions, where we compared $N=10, 20, 30$ to estimate the size effects (see example in Fig.~\ref{f:M_vs_T}a in the SI). Initial annealing is performed for $5\cdot10^4$ steps at the simulated temperature and statistical sampling of different observables is done afterwards for $10^5$ steps.

To make it easier to discuss and present graphically different interactions in the studied spinels, we define effective magnetic interactions which characterize interactions between whole V$_4$ clusters (assuming frozen intra-cluster, magnetic degrees of freedom), instead of single atoms:
\begin{equation}
    J_\mathrm{eff}^{mn} = \sum\limits_{i\in\{m\}} \sum\limits_{j\in\{n\}} J_{ij}, \hspace{5pt} \vec{D}_\mathrm{eff}^{mn} = \sum\limits_{i\in\{m\}} \sum\limits_{j\in\{n\}} \vec{D}_{ij}.
    \label{e:effective_interactions}
\end{equation}
Here, the summation runs over all sites of cluster $m$ and all sites of another cluster $n$. Such effective parameters imply the assumption that the coupling between four spins within each V$_4$ cluster is significantly stronger than between different clusters, which is confirmed by our calculations, described below, and that during spin dynamics at not too high temperature the spins of the same cluster rotate synchronously. Numerical results obtained in this way are discussed below (Sections IV and V) for GaV$_4$S$_8$. These parameters, however, lead to a different temperature-dependent magnetization (from Monte-Carlo simulations) in the ferromagnetic state compared to the interactions $J_{ij}$ and $\vec{D}_{ij}$ between individual V sites (see example in Fig.~\ref{f:M_vs_T}b in the SI). The nature of this effect will be studied in the future.

\vspace{5pt}
\textbf{Step 3 (micromagnetics).} From the atomistic interaction parameters $J_{ij}$ and $\vec{D}_{ij}$ defining the spin model (\ref{e:Heisenberg_model}) one can go to the continuous limit and obtain the micromagnetic energy which can be used to model the magnetic properties on length scales that range up to hundreds of nanometers. Let us consider the derivation of the micromagnetic energy density $\varepsilon_\mathrm{DM}$ due to the DM interaction (derivation for the Heisenberg exchange is similar). Following the derivations in [\onlinecite{Poluektov2018},\onlinecite{Zhang2017}], one can start from the atomistic DM interactions between spin on site $i$ and all other spins on sites $j$:
\begin{equation}
    \varepsilon_\mathrm{DM} = -\sum\limits_{j} \vec{D}_{ij} \cdot (\vec{S}_i \times \vec{S}_j)
    \label{e:DMI_atomistic}
\end{equation}
and replace $\vec{S}_i$ with the micromagnetic order parameter, $\vec{m}\equiv\vec{m}(\vec{r})$, as well as $\vec{S}_j$ with 1$^\mathrm{st}$-order expansion $\vec{m}+(\vec{R}_{ij}\cdot\vec{\nabla})\vec{m}$, where $\vec{R}_{ij}$ is the distance between the two sites. Substituting this into 
Eqn.~(\ref{e:DMI_atomistic}) leads to a DM contribution to the micromagnetic energy density:
\begin{align}
    \varepsilon_\mathrm{DM} &= -\sum\limits_{j} \vec{D}_{ij} \cdot (\vec{m}\times (\vec{R}_{ij}\cdot\vec{\nabla})\vec{m}) =\\
    &= +\vec{m}\cdot \left[ \sum\limits_{j}\vec{D}_{ij} (\vec{R}_{ij}\cdot\vec{\nabla})\right]\times \vec{m},
    \label{e:micromagnetic_derivation}
\end{align}
where, one can define the spiralization matrix $\hat{D} \equiv D_{\alpha\beta}\:(\alpha,\beta=x,y,z)$ as
\begin{equation}
  D_{\alpha\beta} = \sum_{j\neq i} D_{ij}^\alpha R_{ij}^\beta.
  \label{e:spiralization_matrix}
\end{equation}

In general, this matrix can contain nine non-zero components and has to be consistent with the crystal symmetry. For GaV$_4$S$_8$, we find that the only non-zero components are $D_{xy} = -D_{yx} = D$. Due to this specific form, the $x$-component of the vector in the square brackets in Eqn.~(\ref{e:micromagnetic_derivation}) is $D_{xy}\,\partial/\partial y$ and the $y$-component is $-D_{xy}\,\partial/\partial x$, while the $z$-component is zero. Accordingly, the DM contribution to the micromagnetic energy reads:
\begin{equation}
    \varepsilon_\mathrm{DM} = (m_x, m_y, m_z)\cdot\left|
      \begin{array}{ccc}
          \vec{e}_x & \vec{e}_y & \vec{e}_z \\[3pt]
          D\,\cfrac{\partial}{\partial y} & -D\,\cfrac{\partial}{\partial x} & 0 \\[3pt]
          m_x & m_y & m_z
      \end{array}
    \right|.
    \label{e:determinant}
\end{equation}
A straightforward calculation gives the following energy density, $\varepsilon_\mathrm{DM}$, in a form which coincides with the interfacial type of DM interaction often discussed in the literature for magnetic films (see Eqn.~(8) in Ref.~\onlinecite{Bogdanov2001}):
\begin{equation}
    -D \left[ m_x \frac{\partial m_z}{\partial x} - m_z \frac{\partial m_x}{\partial x} + m_y \frac{\partial m_z}{\partial y} - m_z \frac{\partial m_y}{\partial y} \right].
    \label{e:microenergy}
\end{equation}
This result agrees with the Lifshitz invariants expected for the $C_{3\nu}$ crystal symmetry, as discussed, for example, in Ref.~[\onlinecite{Bogdanov2002}] (Eqn.~6 in this reference) and Ref.~[\onlinecite{Ado2020}] (Table~I in this reference), and with the derivation for another spinel GaV$_4$Se$_8$ in the SI of Ref.[\onlinecite{Zhang2017}]. In the latter reference, however, only nearest neighbors and cluster-cluster interactions were taken into account, while in our work we include the full information on the intersite interaction parameters $J_{ij}$ and $\vec{D}_{ij}$ between several hundred and a couple of thousand neighbors.

As a side remark, for bulk systems with cubic crystal symmetry and isotropic DMI, such as B20 compounds MnSi and FeGe, the second line in the determinant in Eqn.~(\ref{e:determinant}) would be $D\cdot(\partial/\partial x,\partial/\partial y,\partial/\partial z)$, since the spiralization matrix $\hat{D}$ is diagonal, as verified by our direct calculations (not shown here). This would lead to the usual expression for the isotropic DM energy density $\varepsilon_\mathrm{DM} = D\,\vec{m}\cdot(\vec{\nabla}\times\vec{m})$. Calculation of the micromagnetic parameters and magnetic textures (helical states and skyrmions) for B20 systems is discussed, for example, in our recent works [\onlinecite{Borisov2021},\onlinecite{Borisov2022}] as well as in Refs.[\onlinecite{Gayles2015,Kashin2018,Grytsiuk2019}].

Similarly, one can also derive the magnetic energy for the Heisenberg exchange:
\begin{align}
    \varepsilon_H &= \sum\limits_{j\neq i} J_{ij} \vec{S}_i \cdot \vec{S}_j \rightarrow\\
    & \sum\limits_{j\neq i} J_{ij} \vec{m}\cdot \left( \vec{m} + (\vec{R}_{ij}\cdot\nabla)\vec{m} + \frac12 (\vec{R}_{ij}\cdot\nabla)^2\vec{m} \right).
    \label{e:Heisenberg_micromagnetic}
\end{align}
Here, the first term is a constant energy contribution, since it is proportional to $\vec{m}^2$, that is equal to 1 (because constant length of magnetic moment vectors is considered). One can also show that the next term with a 1$^{\mathrm{st}}$-order derivative reads:
\begin{equation}
    \vec{m}\cdot (\vec{R}_{ij}\cdot\nabla)\vec{m} \equiv m_\alpha R_{ij}^\beta \nabla_\beta m_\alpha = R_{ij}^\beta \nabla_\beta \left( m_\alpha^2/2 \right).
\end{equation}
which is zero, because $m_\alpha m_\alpha = 1$. Finally, the leading-order term in Eqn.~(\ref{e:Heisenberg_micromagnetic}) reads:
\begin{align}
    \varepsilon_H &= \frac12 \sum\limits_{j\neq i} J_{ij} \vec{m}\cdot (\vec{R}_{ij}\cdot\nabla)^2\vec{m} =\\
    &= \frac12 \sum\limits_{j\neq i} J_{ij} R_{ij}^\alpha R_{ij}^\beta \, m_\gamma \nabla_\alpha \nabla_\beta m_\gamma.
\end{align}

Usually, only the diagonal term is considered in micromagnetic simulations ($\alpha = \beta$), and the energy becomes proportional to the spin stiffness $A$ defined as follows:
\begin{equation}
  A = \frac{1}{2} \sum_{j\neq i} J_{ij} R^2_{ij}.
  \label{e:spin_stiffness}
\end{equation}

As discussed in the literature [\onlinecite{Pajda2001}], and our recent work [\onlinecite{Borisov2022}], the numerical evaluation of micromagnetic parameters according to Eqns.~(\ref{e:spiralization_matrix}) and (\ref{e:spin_stiffness}) shows a convergence problem with respect to the real-space cutoff. For that reason, following the literature recipe in Ref.~[\onlinecite{Pajda2001}], we introduce an exponential regularization factor (in contrast to Eqn.~(\ref{e:effective_interactions}), indices $i$ and $j$ refer either to atomic sites or different clusters):
\begin{equation}
  A = \frac{1}{2}\sum_{j\neq i} J_{ij}R^2_{ij}\,e^{-\mu R_{ij}}, \hspace{2pt} D_{\alpha\beta} = \sum_{j\neq i} D_{ij}^\alpha R_{ij}^\beta \,e^{-\mu R_{ij}}.
  \label{e:micromagnetic_parameters}
\end{equation}
where the limit $\mu \rightarrow 0$ is taken at the final step. The exponential factors are introduced here to improve the convergence with respect to the real-space cutoff for the magnetic interactions. The calculated values of $A$ and $\hat{D}$ as functions of parameter $1.0 < \mu < 2.0$ are then extrapolated to $\mu = 0$ using a 3$^\mathrm{rd}$-order polynomial, which gives a reasonable fitting quality, and a discussion of some further technical details can be found in the SI of Ref.~\onlinecite{Borisov2022}.

As mentioned on page 2, we compare the properties of GaV$_4$S$_8$ for two electronic configurations of V$_4$ clusters which we find to be relatively close in energy ($\Delta E \sim \unit[100]{meV}$). In case of the distributed-moment configuration (Fig.~\ref{f:spinel_structure}b), we calculate the spin stiffness $A$ and DM spiralization $\hat{D}$ from the effective interactions defined by 
Eqn.~(\ref{e:effective_interactions}), because the spin stiffness from the original $J_{ij}$ interactions between individual V sites is negative. This is natural to expect since the cluster configuration is ferrimagnetic in our calculations. The effective cluster-cluster DM interaction, on the other hand, disregards the internal DMI between V sites in the same cluster, which should not matter for large-scale magnetic textures. This is because we assume that the internal magnetic structure of V$_4$ clusters is frozen, which is a good approximation due to the large calculated intracluster magnetic exchange which can have a magnitude as large as $\unit[28]{meV}$. For the localized-moment state (Fig.~\ref{f:spinel_structure}c), we compute the micromagnetic parameters calculated from the original interatomic interactions while taking into account only the V$_1$ sites with the largest magnetic moment. The resulting micromagnetic parameters are scaled down by the square of the V$_1$ moment to simulate an effective $\unit[1]{\mu_\mathrm{B}}$ V$_4$ cluster.

Regarding the DM spiralization matrix (Eqn.~\ref{e:micromagnetic_parameters}), initially we obtain it in the basis of Cartesian unit vectors $\vec{e}_\alpha$ ($xyz$-basis). It turns out, however, that it is more convenient to work in the basis where the $\vec{e}_z$ is along the $[111]$ direction (relevant for skyrmions) and the other two basis vectors are in the $(111)$-plane. In particular, we choose $\vec{e}_y$ as a vector connecting the $[111]$ line with the end of the lattice vector $\vec{a}_2$ and $\vec{e}_x$ is obtained as a normalized cross product of $\vec{e}_y$ and $\vec{e}_z$. The inversion of the matrix, where rows are formed by these new unit vectors $\vec{e}_\alpha$, results in a unitary matrix $\hat{U}$ that defines a transformation to the $[111]$-basis. For the DM spiralization matrix $\hat{D}$ in this new basis, one gets:
\begin{equation}
    \hat{D} = \hat{U}^\mathrm{T} \hat{D}' \,\hat{U},
\end{equation}
where $\hat{D}'$ is the matrix in the $xyz$-basis.

It is in the $[111]$-basis where the spiralization matrix of lacunar spinel GaV$_4$S$_8$ has a dominant component $D_{xy} = -D_{yx} = D$, in agreement with the $C_{3\nu}$ crystal symmetry. For that reason, in the paper we discuss the DM interaction and perform the micromagnetic simulations in the basis where the $z$ axis is along the [111]-direction.

Regarding the on-site anisotropy, we assume in our simulations that the uniaxial anisotropy energy constant $K_1$ is between $\unit[(10-16)]{kJ/m^3}$, as suggested in previous works [\onlinecite{Ehlers2016},\onlinecite{Padmanabhan2019}]. This anisotropy constant is even smaller than $K_1 = \unit[45]{kJ/m^3}$ for \textit{hcp} Co and cubic anisotropy constant $K'_1 = \unit[48]{kJ/m^3}$ for \textit{bcc} Fe but larger than $K_1 = \unit[-0.5]{kJ/m^3}$ for \textit{fcc} Ni. Despite the small value of $K_1$ for GaV$_4$S$_8$, as we will see in the following, it is important to include anisotropy in simulations of magnetic skyrmions in this class of systems.

Using the calculated micromagnetic parameters, including the DM interaction (Eqn.\ref{e:microenergy}) and the literature values of the uniaxial anisotropy $\unit[(10-16)]{kJ/m^3}$, we perform micromagnetic simulations of magnetic textures at finite temperature and in external magnetic field using the \textsc{UppASD\-} [\onlinecite{uppasd},\onlinecite{Eriksson2017}] and \textsc{MuMax3} [\onlinecite{mumax3}] codes. The temperature is varied between 0 and $\unit[30]{K}$ and the external field~-- between 0 and $\unit[600]{mT}$, which corresponds to the experimentally studied range of parameters. The micromagnetic region is described by a $(512\times 512\times 1)$ mesh with an equidistant step $\Delta h$ in all directions ($\Delta h = \unit[0.5]{nm}$ for simulating skyrmions and $\Delta h = \unit[1.0]{nm}$ for simulating ferromagnetic state and spin spirals) and non-periodic boundary conditions. We have also verified the effect of dimension in the $z$-direction and the boundary conditions (see discussion in Section~VI). The magnetization dynamics is described by the Landau-Lifshitz-Gilbert equation [\onlinecite{Landau1935},\onlinecite{Gilbert2004}]:
\begin{equation}
    \frac{\partial \vec{m}_i}{\partial t} = -\frac{\gamma}{1 + \alpha^2} \left( \vec{m}_i \times \vec{B}_i + \frac{\alpha}{m}\,\vec{m}_i \times (\vec{m}_i \times \vec{B}_i) \right),
    \label{e:LLG_equation}
\end{equation}
where $\vec{m}_i$ is the magnetization of a given micromagnetic region $i$, and the effective field $\vec{B}_i$ is determined from the micromagnetic parameters $A$, $D$ and uniaxial anisotropy, $K$, as well as the external field, $\vec{B}$, and the classical dipolar field; $\gamma$ is the gyromagnetic ratio. At finite temperature, random field proportional to $\sqrt{\alpha\, T}$ is added to $\vec{B}_i$, and the damping constant $\alpha$ is set to $0.04$. Variable time step in the range $\unit[3\cdot10^{-15}-5\cdot10^{-14}]{s}$ is used for the dynamics simulations, and the total simulation time was around $\unit[(2-5)]{ns}$, which allowed to reach the equilibrium state starting from a random magnetic configuration.

Micromagnetic simulations are run starting from a random magnetic configuration at a temperature of $\unit[20]{K}$. The system is cooled in $\unit[2]{K}$-temperature steps down to $\unit[0]{K}$, which corresponds to simulated annealing, and then the system is heated up to $\unit[20]{K}$ with the same speed. Results for the anisotropy values $K = \unit[10]{kJ/m^3}$ and $K = \unit[16]{kJ/m^3}$ are compared. Since the [111]-axis of the crystal structure is defined now as the $z$-axis, which leads to expression (\ref{e:microenergy}) for the DMI energy, the anisotropy easy-axis is also along the $z$-direction.

\medskip
\textbf{Data Availability} \par
The datasets used and/or analysed during the current study are available from the corresponding author on reasonable request.

\section{Electronic properties}

In the following, we discuss the electronic structure, and related properties, of the lacunar spinel GaV$_4$S$_8$, calculated using the local-density (LDA) as well as the generalized-gradient (GGA) approximation, also including correlations described on the static mean-field level of spin-polarized DFT with Hubbard-$U$ corrections. It is noteworthy that the electronic structure work presented here shows two stable minima, one with a total moment of $\sim$ 1 $\mu_B$ for each tetrahedron of V atoms of the lacunar spinel structure (Fig.~\ref{f:spinel_structure}c) and one solution in which the total moment per V tetrahedron still is close to 1 $\mu_B$, but where one of the V atoms of each tetrahedron carries a significantly larger moment compared to the other V atoms (Fig.~\ref{f:spinel_structure}b). We refer to the first configuration as the distributed moment case, and the second configuration as the localized moment case.

\begin{figure}
{\centering
\includegraphics[width=0.99\columnwidth]{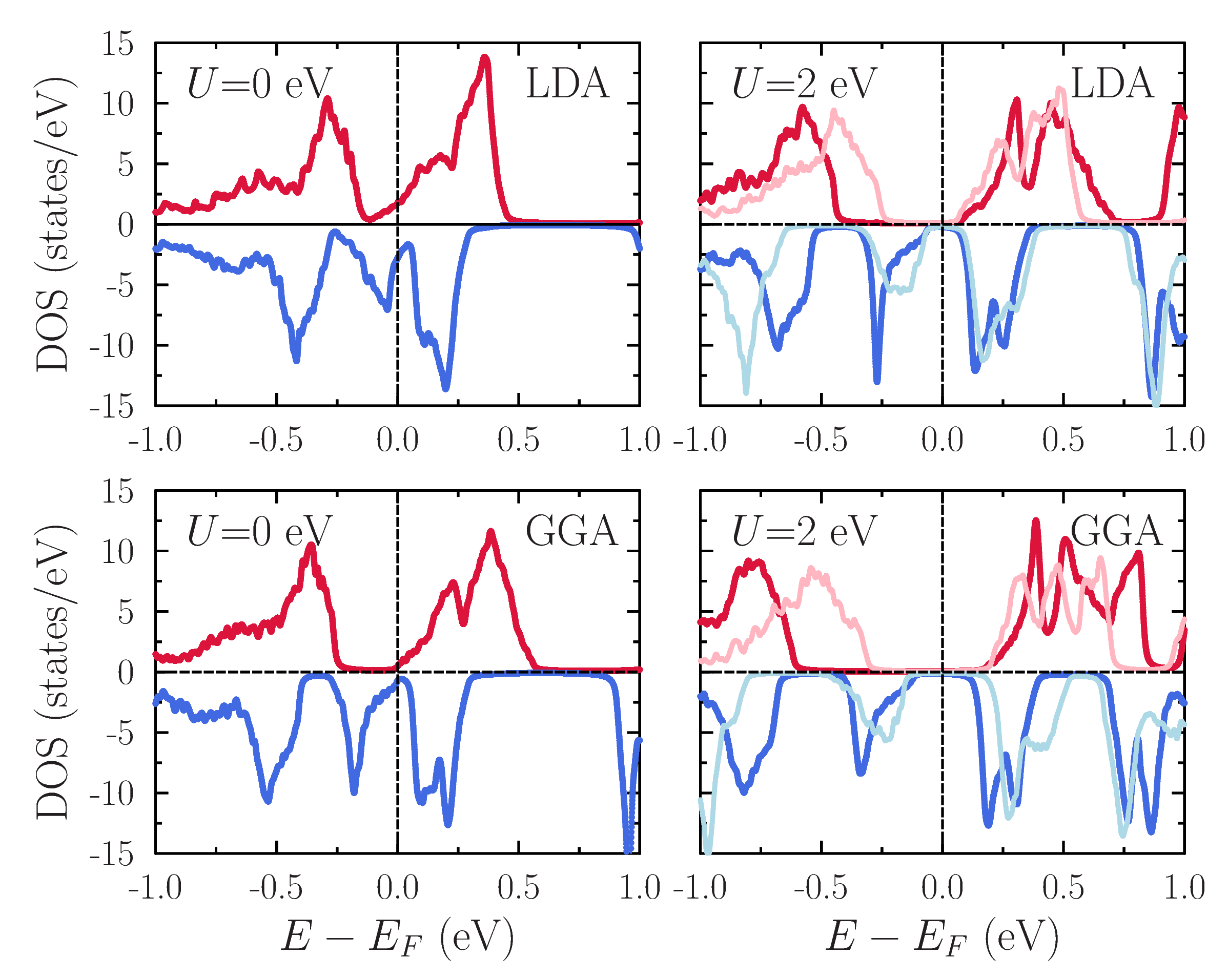}
}
\vspace{-15pt}
\caption{
Density of states of GaV$_4$S$_8$ calculated within density functional theory in LDA and GGA approximations without and with Hubbard-$U$ corrections. Red and blue curves correspond to spin-up and -down states. The system is in the localized-moment semi-metallic state in GGA at $U=\unit[0]{eV}$ (no bands cross $E_\mathrm{F}$, but the energy gap is zero valued) and is semiconducting at $U=\unit[2]{eV}$ with a gap $E_g = \unit[0.2]{eV}$ (localized-moment state, dark lines) and $\unit[0.3]{eV}$ (distributed-moment state, light lines).
}
\label{f:electronic_structure_GaV4S8}
\end{figure}

The calculated electronic states of GaV$_4$S$_8$ near the Fermi level, $E_\mathrm{F}$, are represented by a number of states with a narrow bandwidth crossing $E_\mathrm{F}$, as indicated by the LDA results for the localized moment state of V$_4$ clusters in Fig.~\ref{f:electronic_structure_GaV4S8}a. The spin-splitting of bands increases somewhat in response to electronic correlations in DFT+$U$ calculations and a small gap of the order of $\unit[0.1]{eV}$ appears in the electronic spectrum for $U=\unit[2]{eV}$ (Fig.~\ref{f:electronic_structure_GaV4S8}c). Also, the distributed moment state (Fig.~\ref{f:spinel_structure}b) of V$_4$ clusters becomes stable, but we find that the energy of the localized moment state is somewhat lower by around $\unit[76]{meV/f.u}$. Varying $U$ between 0 and $\unit[2]{eV}$ increases the total magnetic moment per formula unit from $\unit[0.8]{\mu_\mathrm{B}}$ to almost $\unit[1.0]{\mu_\mathrm{B}}$, and these values agree with the range of measured values in the literature [\onlinecite{Pocha2000}].

The results discussed above are obtained within the local-density approximation. Using the generalized-gradient approximation (GGA) we obtain essentially the same trends, but GGA shows a stronger tendency to magnetism. This leads to the magnetic moment of each V$_4$ tetrahedron being close to $\unit[1.0]{\mu_\mathrm{B}}$ already for pure DFT  ($U=\unit[0]{eV}$) calculations (Table~I). At increasing correlation strength $U$, the moments of both V$_1$ and V$_2$ sites are enhanced (Table~I), but the total moment of each V$_4$ cluster remains almost the same. Next, we notice that the DFT+$U$ band structures and densities of states (Fig.~\ref{f:electronic_structure_GaV4S8}) within LDA and GGA are similar but there is an offset around $\unit[1]{eV}$ in terms of the $U$ values between the two approximations, meaning that in GGA+$U$ one needs $U$ values smaller by roughly $\unit[1.0]{eV}$ to get results similar to LDA+$U$. At $U=\unit[2]{eV}$, we find again that the distributed-moment state can be stabilized (last row in Table~I) but it is higher in energy by $\unit[99]{meV/f.u.}$ compared to the localized-moment state. Finally, the band gap calculated in GGA is larger compared to the LDA estimate, which is in accordance with literature (see Fig.\,10 in [\onlinecite{YiqunWang2019}]).

\setlength{\tabcolsep}{6pt}
\renewcommand{\arraystretch}{1.5}
\begin{table}[h]
 \caption{Calculated magnetic moments of V$_1$ and V$_2$ sites and total moment for GaV$_4$S$_8$ with experimental structure. Results are obtained within GGA+$U$ for the localized-moment (``L'' state) and distributed-moment (``D'' state) configurations (Fig.~\ref{f:spinel_structure}b,c).}
\vspace{5pt}
  \centering
  \begin{tabular}{c|c|ccc}
    \hline
    $U$ (eV) & state & V$_1$ & V$_2$ & $\sum m_i$ ($\mu_\mathrm{B}$/f.u.) \\
    \hline
    0.0 & ``L'' & 1.04 & -0.05 & 0.97 \\
    0.5 & ``L'' & 1.22 & -0.11 & 0.98 \\
    1.0 & ``L'' & 1.34 & -0.15 & 0.99 \\
    1.5 & ``L'' & 1.44 & $\approx-0.2$ & 0.98 \\
    2.0 & ``L'' & 1.52 & -0.22 & 0.99 \\
    2.0 & ``D'' & 0.71 & -0.58 & 1.00 \\
    \hline
  \end{tabular}
  \label{t:magnetic_moments}
\end{table}

\section{Magnetic exchange}

At $U=\unit[0]{eV}$, both in LDA and in GGA, we find a considerable ferromagnetic interlayer Heisenberg interaction ($J_\perp$) of the order of $\unit[4]{K}$, which couples the neighboring (111)-planes of V$_4$ clusters. Furthermore, we find a much weaker intralayer interaction ($J_{||}$), which is between the neighboring clusters in each of these (111)-planes. This picture changes when correlations are included within DFT+$U$. In particular, the intralayer exchange $J_{||}$ becomes stronger (reaching up to almost $\unit[12]{K}$) and can outweigh the interlayer exchange $J_\perp$. Interestingly, within the LDA, $J_\perp$ increases as a function of $U$ until $U=\unit[1]{eV}$ and then decreases, while GGA results show decreasing $J_\perp$ which even becomes antiferromagnetic at $U=\unit[2]{eV}$ where the localized-moment state of the V$_4$ clusters changes to the distributed-moment state (Fig.~\ref{f:spinel_structure}c). In LDA, however, such a transformation of the V$_4$ cluster does not lead to change of sign of effective magnetic interactions.

It is necessary to emphasize that the parameters $J_\perp$ and $J_{||}$ that we discuss here are actually the effective interactions defined by Eqn.~(\ref{e:effective_interactions}) and are introduced to make the discussion of the results more transparent. For Monte Carlo (MC) simulations at different temperatures, we use the original data, i.e. interactions between individual V sites (even weakly magnetic ones). One should note that, even though the effective interlayer exchange becomes AFM for $U=\unit[2]{eV}$ in GGA, the Monte Carlo simulations using not just effective but all calculated intersite interactions actually find the correct ferromagnetic ground state, allowing to conclude that the effective cluster-cluster models probably cannot cover the whole physics in this system (see further Monte Carlo simulations in Fig.~\ref{f:M_vs_T}b in the SI).

\begin{figure}
{\centering
\includegraphics[width=0.95\columnwidth]{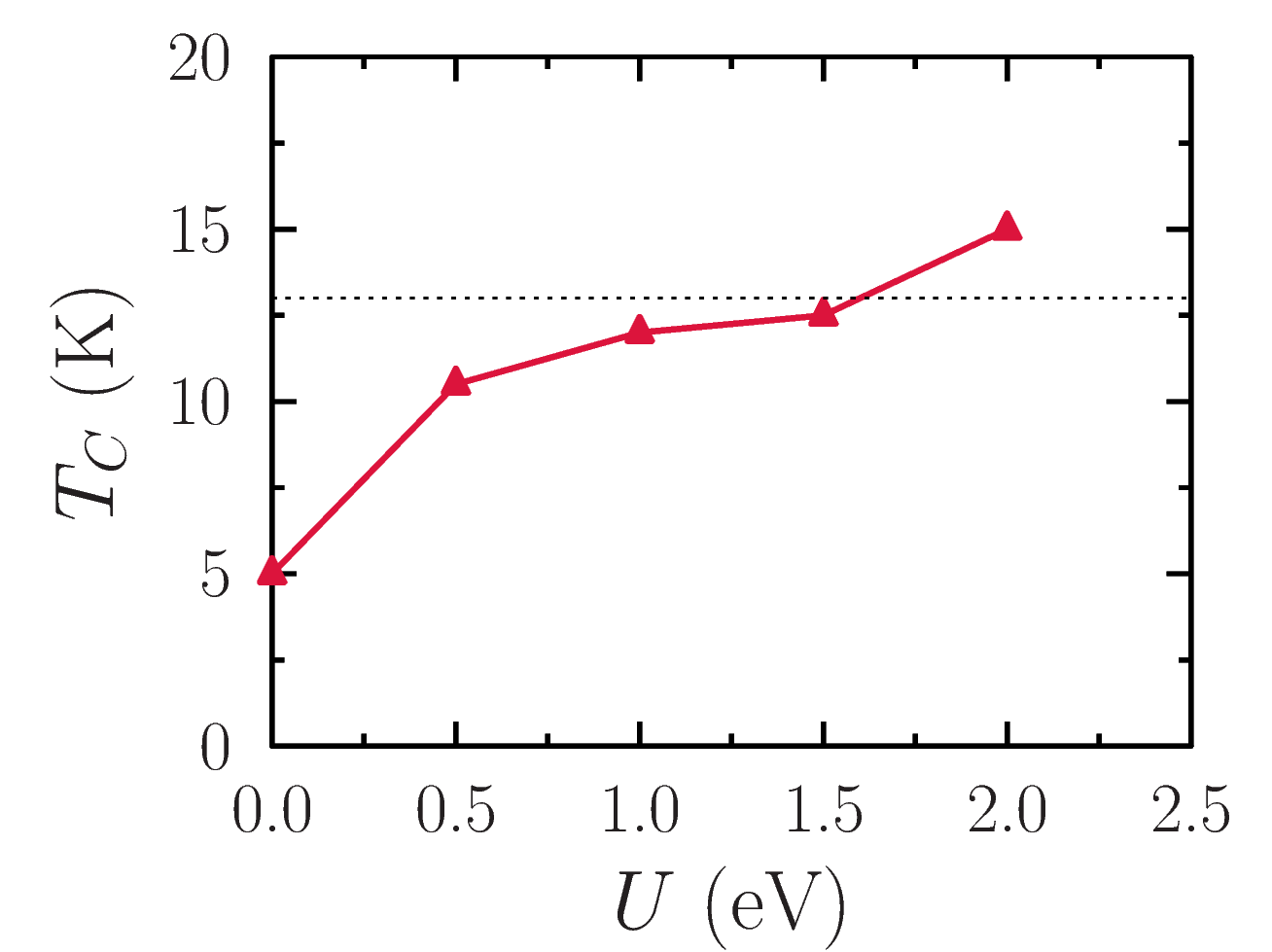}
}
\vspace{-5pt}
\caption{Curie temperature estimated from classical atomistic Monte Carlo simulations based on first-principles Heisenberg $J_{ij}$ and Dzyaloshinskii-Moriya $\vec{D}_{ij}$ intersite interactions calculated within GGA+$U$ with variable correlation strength $U$. The dotted line represents the experimental Curie temperature.
}
\label{f:Curie_temperature}
\end{figure}
\begin{figure}
{\centering
\includegraphics[width=0.95\columnwidth]{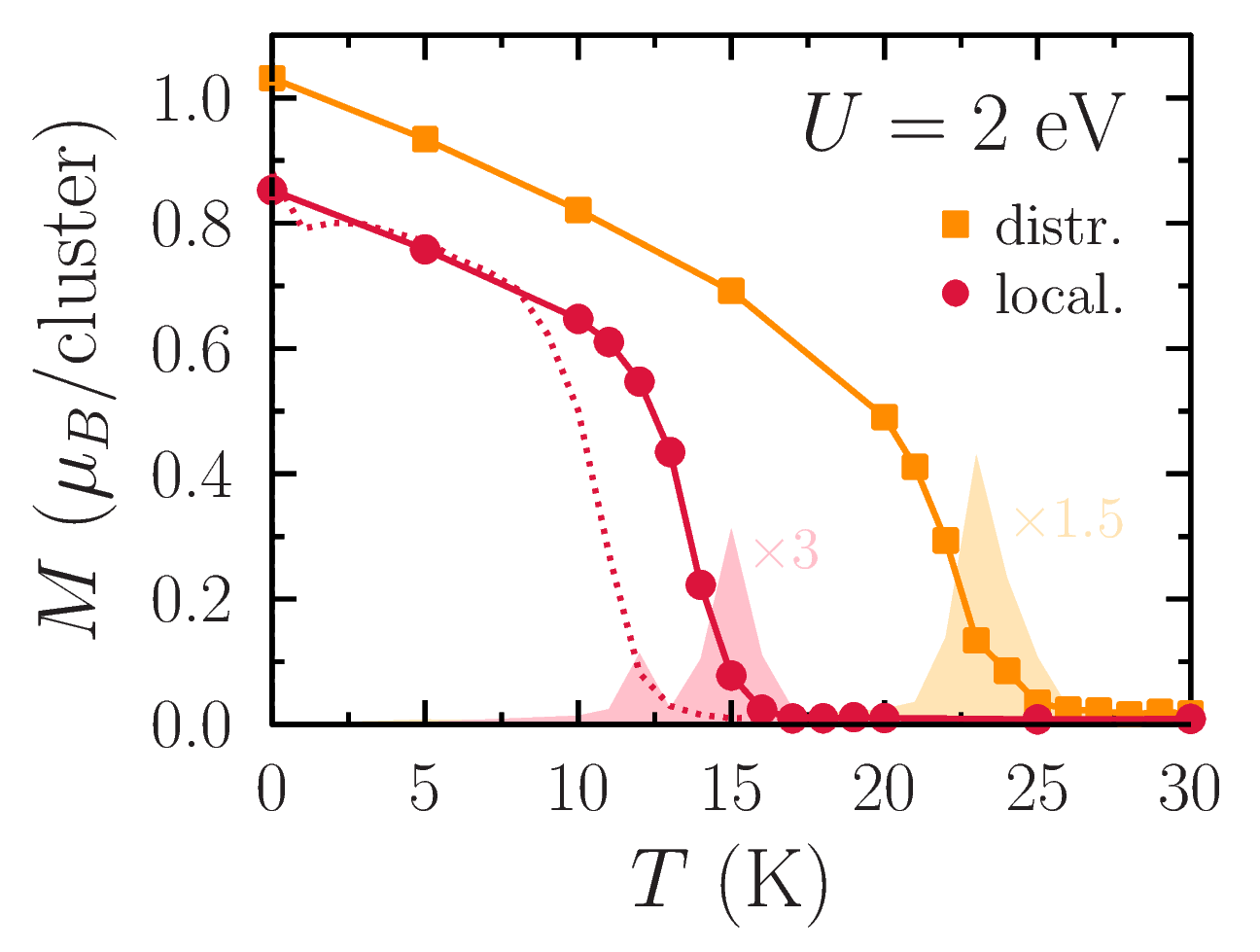}
}
\vspace{-10pt}
\caption{Temperature-dependence of the magnetization $M(T)$ from Monte Carlo simulations based on magnetic interactions at $U=\unit[2]{eV}$ (solid lines), where both the localized and distributed moment states can be stabilized and compared. $(30\times30\times30)$ supercell was used. The dotted line shows the $M(T)$ for the localized moment state at $U=\unit[1]{eV}$. The shaded regions correspond to the magnetic susceptibility, the peak of which indicates the Curie temperature.}
\label{f:M_vs_T_U=2}
\end{figure}

Figure~\ref{f:Curie_temperature} shows how the Curie temperature, estimated from our MC simulations using both the calculated Heisenberg and Dzyaloshinskii-Moriya interactions, changes as a function of electronic correlation strength characterized by the $U$ parameter. For the localized-moment state, the ordering temperature for $U=\unit[(1.0-1.5)]{eV}$ is around $\unit[12]{K}$ and agrees nicely with the measured value of $\unit[13]{K}$ [\onlinecite{Kezsmarki2015}]. Electronic correlations increase slightly the Curie temperature (by a few degrees) and at some point ($U=\unit[2]{eV}$) allow to stabilize the distributed moment state, as discussed above. The temperature dependence of magnetization ($M(T)$) curves for both cluster configurations is similar (Fig.~\ref{f:M_vs_T_U=2}) and look like typical $M(T)$ curves for ferromagnets. However, the distributed-moment state is characterized by a higher Curie temperature (around $\unit[23]{K}$), which is overestimated compared to the experimental value of $\unit[13]{K}$. Stronger magnetism for the distributed-moment configuration is likely caused by the larger number of magnetic exchange paths when all V sites in each V$_4$ cluster have non-zero magnetic moments.

\section{Dzyaloshinskii-Moriya interaction}

For GaV$_4$S$_8$ within pure DFT (LDA approach, $U=\unit[0]{eV}$), we obtain the spiralization matrix in the [111]-basis (in units of meV$\cdot$\AA)\footnote{In this paper, all DM spiralization matrices are given with a precision of two digits after comma and all numbers with the absolute value below 0.01 are written as 0.}, as described in the Methods section:
\begin{equation}
    \hat{D} = \left(
      \begin{array}{ccc}
         0 &  +0.12 & 0 \\ 
        -0.12 &  0 & 0 \\
         0 &  0 & 0 \\
      \end{array}
    \right).
\end{equation}
The $[111]$-basis appears to be more convenient for studying the micromagnetic behavior than the Cartesian basis, since the $\hat{D}$ matrix has just one dominating component $D_{xy}$ which is to be substituted as the $D$ parameter in Eqn.~(\ref{e:microenergy}). The obtained form of the spiralization matrix agrees with a previous work [\onlinecite{Zhang2017}] and the observation of N\'eel skyrmions in this bulk system [\onlinecite{Kezsmarki2015}]. Possible origin of the slight asymmetry ($\sim 10^{-3}$) of the calculated spiralization matrix may be related to a weak dependence of electronic properties on the total magnetization direction due to spin-orbit coupling.

\begin{figure}
\vspace{5pt}
{\centering
\includegraphics[width=0.99\columnwidth]{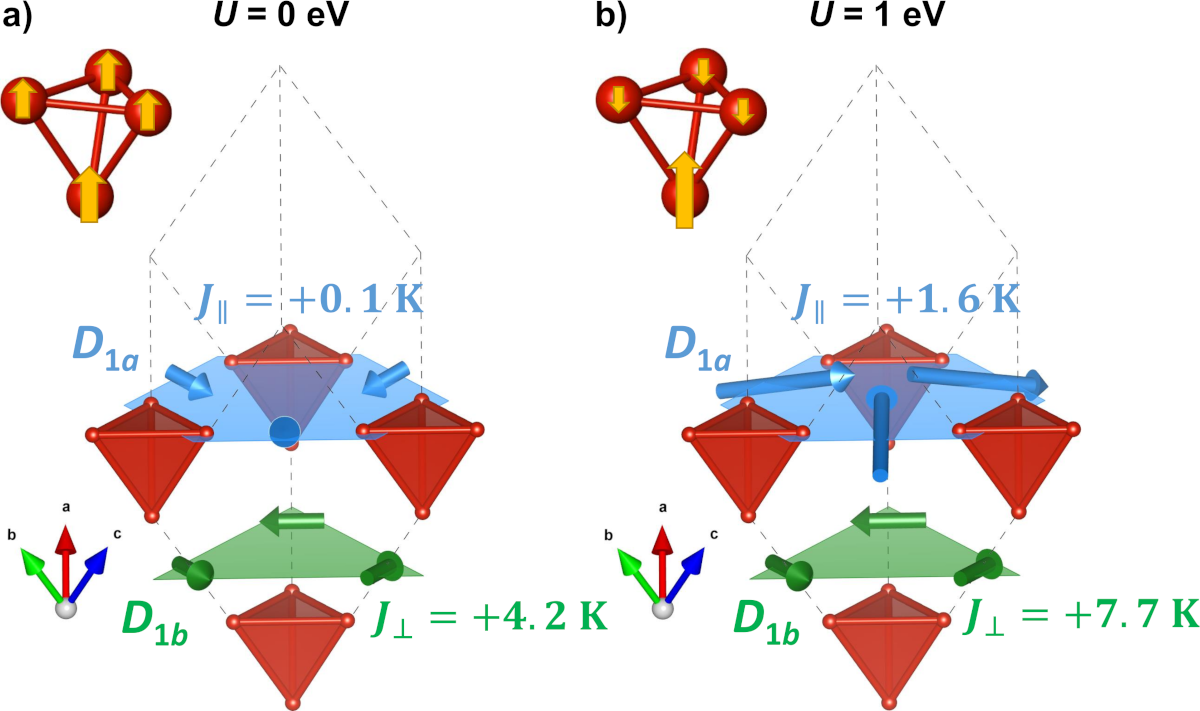}
}
\vspace{-15pt}
\caption{Schematic representation of the effective (cluster-cluster) DM vectors for nearest neighbors in GaV$_4$S$_8$ by arrows, where the arrow length is proportional to the DMI magnitude. The LSDA and LSDA+$U$ results are shown in comparison. In both cases, the DM vectors follow the $C_{3\nu}$ crystal symmetry. For the same bonds, the effective Heisenberg exchange (\ref{e:effective_interactions}) is given in Kelvin (positive value indicates a ferromagnetic interaction).}
\label{f:DM_interactions}
\end{figure}

When electronic correlations are included on the mean-field level with LDA and $U=\unit[1]{eV}$, the DMI increases dramatically by more than a factor of 4 and changes sign:
\begin{equation}
    \hat{D} = \left(
      \begin{array}{ccc}
          0    & -0.54  &  0 \\
          0.54 &  0     &  0 \\
          0    &  0     &  0 \\
      \end{array}
    \right)
\end{equation}
Similar response to moderate correlations is observed for the cluster-cluster DM interactions, in particular, for the nearest-neighbor in-plane interaction $D_{1a}$ (Fig.~\ref{f:DM_interactions}). Large enhancement of the DMI can be related to the change of the magnetic state of the V$_4$ clusters: at $U=\unit[1]{eV}$ the magnetic moment of one of V sites increases by more than a factor of 2, while the other three sites remain weakly magnetic but change to the opposite direction. This more asymmetric distribution of magnetization at $U=\unit[1]{eV}$ can, in principle, create additional inversion-symmetry breaking effect which increases the DMI. On top of that, the system becomes semimetallic (Fig.~\ref{f:electronic_structure_GaV4S8}), since the band gap is zero but no bands directly cross the Fermi level.

Within the generalized-gradient approximation (GGA), we find a contrasting behavior, since the DMI parameter $D_{xy}$ decreases in absolute value for the localized-moment state as a function of electronic correlations $U$, from $D_{xy}=\unit[-0.20]{meV\cdot\AA}$ at $U=\unit[0]{eV}$ to $\unit[-0.16]{meV\cdot\AA}$ at $U=\unit[2]{eV}$. We should mention that, in contrast to LDA, correlations added within GGA lead to a band gap opening already at $U=\unit[1]{eV}$ (Fig.~\ref{f:electronic_structure_GaV4S8}) which could partially explain the differences in the calculated DMI. Notably, for the distributed-moment state in GGA, which is also stable at $U=\unit[2]{eV}$, the DM parameter $D_{xy}=\unit[-1.37]{meV\cdot\AA}$ is considerably larger. Strong dependence of the DM spiralization constant on the cluster state makes sense in view of the other findings shown in Fig.~\ref{f:DM_interactions} (for LDA approximation) where the atomistic DMI is larger for the more asymmetric magnetization distribution in the cluster which may break further the inversion symmetry. Also, the localized- and distributed-moment states of V$_4$ cluster lead to distinct band structures (Fig.~\ref{f:electronic_bands_GaV4S8} in the SI) and this can have a sizeable effect on the DMI too.

In addition, the spin stiffness $A$ decreases by almost a factor of 5 (in GGA) in response to additional electronic correlations with $U=\unit[2]{eV}$. Because of that, the $A/D_{xy}$ ratio decreases from $\unit[26.3]{nm}$ to $\unit[7.4]{nm}$, suggesting again that the magnetic properties of GaV$_4$S$_8$ are very sensitive to electronic correlations. On the other hand, at $U=\unit[2]{eV}$ also the distributed moment state is stable and it shows a relatively large spin stiffness, compared to the localized moment state. Also, the DM interaction is considerably larger resulting in the $A/D$ ratio of $\unit[5.0]{nm}$. The $A/D$ ratio is important in the context of non-collinear magnetism as well as skyrmions and lower values of $A/D$, in general, are expected to indicate more compact skyrmions and helical spin states, which is confirmed by the micromagnetic simulations in Sec.~VI. Notably, the LDA+$U$ results show a spin stiffness which is roughly a factor of two larger compared to the GGA+$U$ estimates for $U\geq\unit[1]{eV}$.

It is worth mentioning that the strongest interatomic DM interaction ($D_0\sim\unit[0.6]{meV}\approx\unit[6.8]{K}$) in these spinels, according to our calculations, comes from the V-V bonds within each metal cluster, implying that the actual magnetic state of V$_4$ clusters may be non-collinear. On the other hand, the non-collinearity is not expected to be large, since the canting angles within the cluster should be of the order of $D_0/J_0$ where $J_0$ is the nearest-neighbor Heisenberg V-V exchange which is, as we find, antiferromagnetic and in the range of several hundred Kelvin. For that reason, canting angles around a few degrees can be expected, which should not change the main findings reported in the present work.

This intracluster DMI would actually contribute significantly to the calculated micromagnetic constant $D$, and it is a subtle question whether to include this DMI or not when addressing the behavior of skyrmions in this system. A strong argument not to do so, we suggest, is that the internal magnetic exchange in each V$_4$ cluster is very large in our calculations and, for that reason, the spins of each cluster are expected to co-rotate. In that case, the internal DMI as well as the internal Heisenberg exchange do not contribute to the micromagnetic energy.
Table~II summarizes our findings for the spin stiffness and DM spiralization for the different calculation setups and two V$_4$ cluster states.

\setlength{\tabcolsep}{6pt}
\renewcommand{\arraystretch}{1.5}
\begin{table}[h]
 \caption{Calculated micromagnetic parameters (spin stiffness $A$ in units of meV$\cdot\mathrm{\AA}^2$ and spiralization $D_{xy}$ in units of meV$\cdot\mathrm{\AA}$) for GaV$_4$S$_8$ with experimental structure. The $A$ and $D_{xy}$ parameters are obtained from V$_1$-V$_1$ (in case of localized-moment state, ``L'') or cluster-cluster (in case of distributed-moment state, ``D'') interactions, which exclude the intracluster exchange. $A$ and $D_{xy}$ for the ``L'' state are divided by the square of the magnetic moment of the V$_1$ site. Estimate of the Curie temperature ($T_\mathrm{C}$) is obtained from classical Monte Carlo simulations using V-V interaction parameters for all V sites. Results obtained within GGA+$U$ for different correlation strengths $U$ and cluster states ``L'' and ``D'' (see main text and Fig.~\ref{f:spinel_structure}b,c) are shown for comparison.}
\vspace{5pt}
  \centering
  \begin{tabular}{c|c|cccc}
    \hline
    $U$ (eV) & state & $A$ & $D_{xy}$ & $A/D$ (nm) & $T_\mathrm{C}$ (K) \\
    \hline
    0.0 & ``L'' & 53.14 & $-0.202$ & 26.3 & $\approx 0$ \\
    0.5 & ``L'' & 35.08 & $-0.200$ & 17.5 &    \\
    1.0 & ``L'' & 23.77 & $-0.180$ & 13.2 & 12 \\
    1.5 & ``L'' & 15.77 & $-0.166$ &  9.5 &    \\
    2.0 & ``L'' & 11.48 & $-0.155$ &  7.4 & 15 \\
    2.0 & ``D'' & 68.40 & $-1.370$ &  5.0 & 23 \\
    \hline
  \end{tabular}
  \label{t:micromagnetic_parameters}
\end{table}

\section{Micromagnetic simulations}

Part of the motivation for this study is the experimental realization of N\'eel skyrmions  in the lacunar spinel GaV$_4$S$_8$. As an ultimate test of the accuracy of the calculated electronic structure and interatomic exchange parameters, as well as the transition to micromagnetic interaction strengths, we explore here the possibility of skyrmion formation with the aforementioned parameters. It should be noted that so far no fitting to experimental data has been made and all calculations are made in {\it ab-initio} mode. To undertake this investigation we compare three sets of calculated micromagnetic parameters and their ability to reproduce skyrmions. The parameters used are:
\begin{enumerate}
    \item[a)] the localized-moment state at $U = \unit[1]{eV}$:\\
    $A = \unit[0.1694]{pJ/m}$, $D_{xy} = \unit[0.0128]{mJ/m^2}$
    \item[b)] the localized-moment state at $U = \unit[2]{eV}$:\\
    $A = \unit[0.0824]{pJ/m}$, $D_{xy} = \unit[-0.0112]{mJ/m^2}$
    \item[c)] the distributed-moment state at $U = \unit[2]{eV}$:\\
    $A = \unit[0.4886]{pJ/m}$, $D_{xy} = \unit[-0.0979]{mJ/m^2}$
\end{enumerate}
Note that the values are different here when compared to those in Table~II, because they are divided by the unit cell volume and converted to the units usually used in experimental reports. The saturation magnetization of the simulations is $\unit[41.35]{kA/m}$ which corresponds to $\unit[1]{\mu_B}$ per formula unit, a value found in experiment as well as in the calculations (Sec.III).

\begin{figure}
{\centering
\includegraphics[width=0.99\columnwidth]{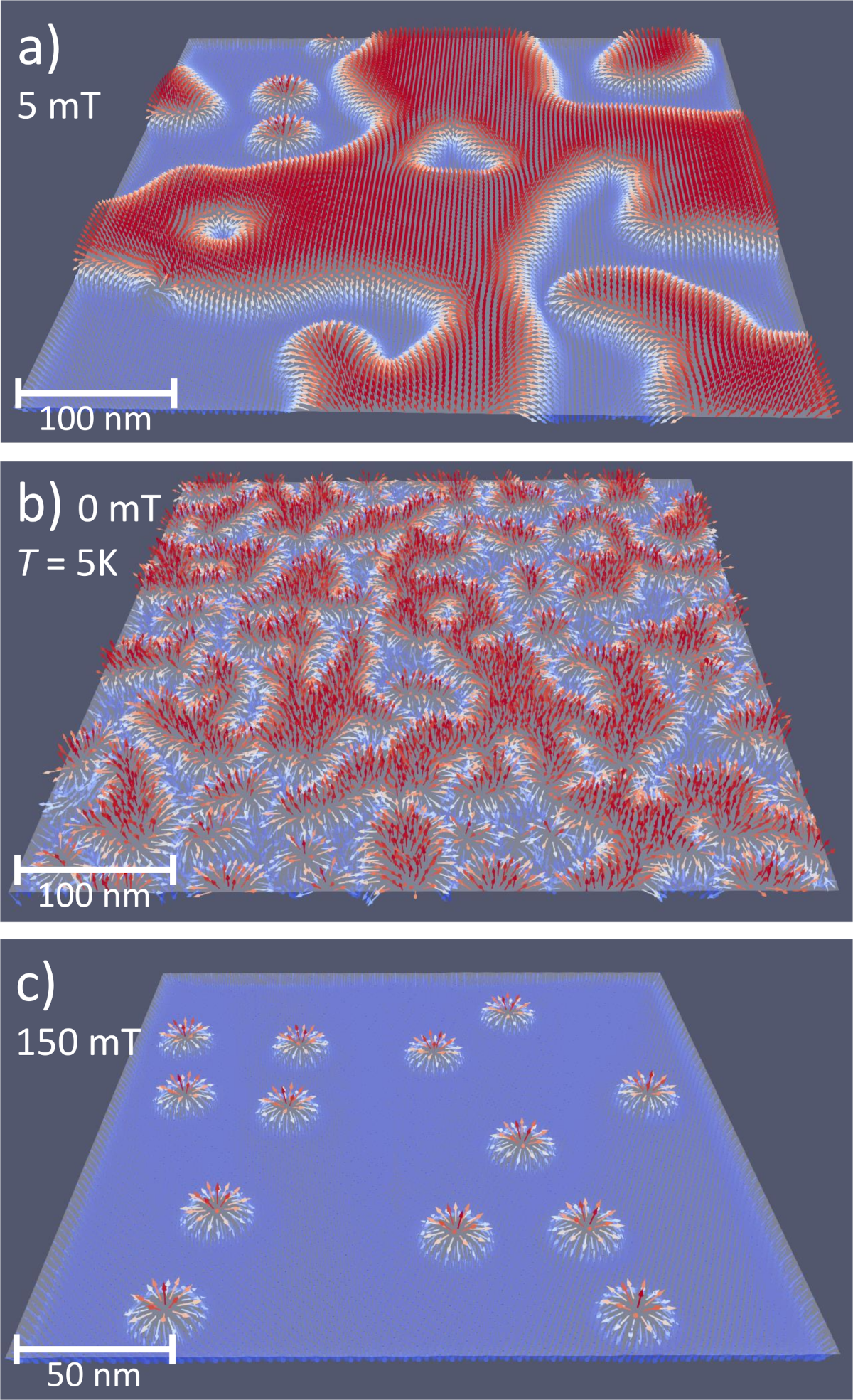}
}
\vspace{-10pt}
\caption{
Micromagnetic simulation results (obtained with \textsc{MuMax3} code) showing the annealed zero-temperature magnetic configurations: a) ferromagnetic domains with isolated skyrmions in external magnetic field $B = \unit[5]{mT}$, b) zero-field spin-spiral state at $T=\unit[5]{K}$, and c) N\'eel skyrmions with a size of $\unit[17]{nm}$ for $B = \unit[150]{mT}$. The size of the simulation region is $\unit[(512\times512\times1)]{nm}$ in a,b) and $\unit[(256\times256\times0.5)]{nm}$ in c). The V$_4$ clusters are in the distributed-moment state and the uniaxial anisotropy is set to $K = \unit[16]{kJ/m^3}$ (other parameters are from the last row of Table~II). The color code shows the out-of-plane direction of the magnetic moment vector in each point of space.
}
\label{f:micromagnetics}
\end{figure}

\begin{figure}
\vspace{10pt}
{\centering
\includegraphics[width=0.99\columnwidth]{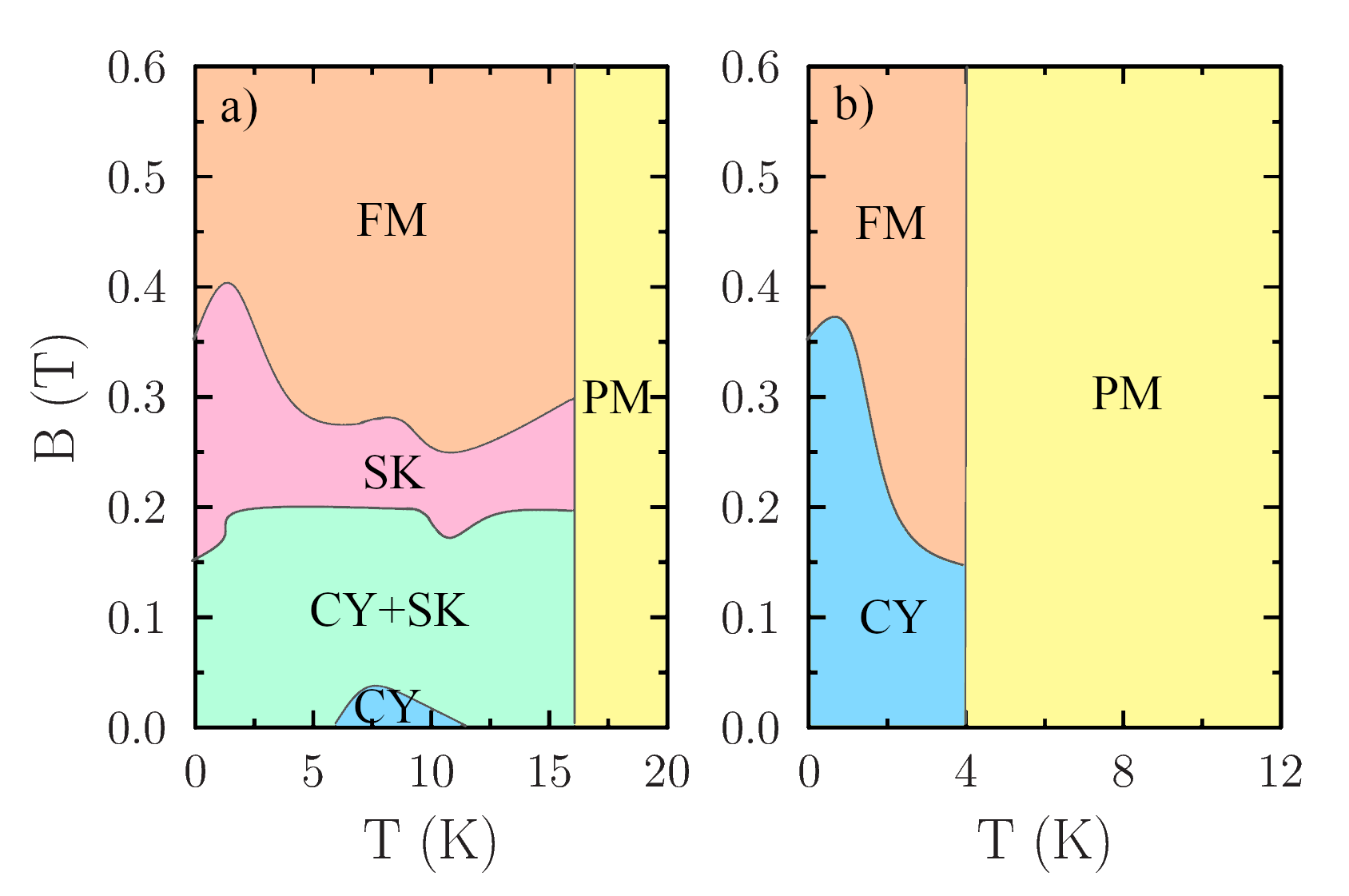}
}
\vspace{-15pt}
\caption{Theoretical phase diagrams of GaV$_4$S$_8$ for the a) distributed- and b) localized-moment V$_4$ cluster configurations based on the micromagnetic simulations with \textsc{MuMax3} code. Temperature and external magnetic field ranges are comparable to the experimentally studied ones. Different phases are shown: ferromagnetic (FM, orange), skyrmionic (SK, pink), cycloid+skyrmions (CY+SK, mint green), cycloid (CY, light blue) and paramagnetic (PM, yellow). The original data points for these phase diagrams are shown in Fig.~\ref{f:phase_diagrams_data} in the SI}
\label{f:phase_diagrams}
\end{figure}

For the parameter set ``c'' (distributed-moment state), we obtain a multi-domain ferromagnetic state with isolated skyrmions (Fig.~\ref{f:micromagnetics}a) at zero temperature and zero applied field. The emergence of a spin-spiral magnetic state (Fig.~\ref{f:micromagnetics}b), for zero-field simulations at low but finite temperature, with a period of around $\unit[20]{nm}$, agrees well with the measured value $a_\mathrm{cyc}=\unit[17.7]{nm}$ (see Fig.~3 in Ref.~\onlinecite{Kezsmarki2015}). In addition, a state with stable skyrmions, with calculated topological charge close to $\pm 1$, is found  when the external magnetic field is applied along the $\pm z$-direction with a strength between $\unit[(50-300)]{mT}$ (see Fig.~\ref{f:micromagnetics}c). The skyrmion size in our simulations depends on the external field and ranges from $\unit[27]{nm}$ at $B = \unit[25]{mT}$ to $\unit[13]{nm}$ at $B = \unit[300]{mT}$, where the number of skyrmions is dramatically decreased. This estimated size is compatible with the experimentally observed skyrmion size $a_\mathrm{sky}=\unit[22.2]{nm}$ reported in Fig.~3 of Ref.~\onlinecite{Kezsmarki2015}. At higher fields, the system is ferromagnetic up to temperatures around $\unit[(12-14)]{K}$. The latter marks the critical temperature also for other types of magnetic order in this system. All these findings are summarized in the calculated phase diagram in Fig.~\ref{f:phase_diagrams}a and are in a good agreement with experiments in Ref.~\onlinecite{multiferroic}.

In contrast, the parameter sets ``a'' and ``b'' (obtained from the localized moment state) produce practically no skyrmions and show a magnetic order only at relatively low temperatures (see phase diagram in Fig.~\ref{f:phase_diagrams}b), which is due to the smaller values of the $A$ and $D$ parameters. In general, larger spin stiffness for the distributed moment state can be explained by the fact that there are more interaction paths in the structure, since all V atoms are then magnetic. By varying the micromagnetic parameters $A$, $D$ and $K$, we find (data not shown) that the $A/D$, $A/K$ and $D/K$ ratios are all important for stabilizing skyrmions, which explains why the parameter set ``c'' (distributed moment state) gives a better agreement with experiment, given that the anisotropy is in the range $\unit[(10-16)]{kJ/m^3}$.

Our conclusions for the three parameter sets used in the micromagnetic simulations are qualitatively robust with respect to at least 10\%-variations of the strength of the interactions. We note that lower DM values lead to smaller saturation fields for stabilizing a ferromagnetic state. Results for non-periodic and periodic boundary conditions in the simulations are also found here to yield similar results. We have also verified that increasing the out-of-plane dimension (parallel to the external magnetic field; the z-axis) of the simulation cell from $n_z = 1$ to $n_z = 16$ does not change qualitatively the phase diagrams shown in Fig.~\ref{f:phase_diagrams} (which are obtained using $n_z = 1$). The most prominent, quantitative change of these simulations, compared to the thin 2D simulation cell, is an increase of the Curie temperature for the distributed moment configuration up to around $\unit[20]{K}$ and a lowering of the magnetic field needed to induce the ferromagnetic state. For the localized-moment configuration, the Curie temperature remains essentially the same for thinner and thicker simulation  cells.

To summarize this section, our results indicate that the distributed-moment configuration (Fig.~\ref{f:spinel_structure}c) describes better the magnetic phase diagram (Fig.~\ref{f:phase_diagrams}a) of bulk GaV$_4$S$_8$ spinel, which shows ferromagnetic phase, spin spirals and N\'eel skyrmions. An example of the magnetic texture of a N\'eel skyrmion is shown in Fig.~\ref{f:neel_skyrmion}, isolated in a ferromagnetic background a) and in a lattice b). Based on the experimental phase diagram reported in Ref. \onlinecite{multiferroic}, the results shown in Fig.~\ref{f:phase_diagrams}a) have similar trends for the transition to the paramagnetic phase. In addition, the transition to the ferromagnetic phase occurs at small magnetic fields (in the range around $\unit[250-400]{mT}$-as shown in Fig.~\ref{f:phase_diagrams}a)) which agrees with observations (in the experiments it is $\unit[50-160]{mT}$). The Curie temperature calculated in our simulations is in the neighborhood of $\unit[17]{K}$ which is close to the experimental one, i.e. $\sim\unit[13]{K}$. The only difference between phase diagrams is related to the skyrmionic and cycloidal regions. In the experimental phase diagram, the skyrmion lattice phase only appears from $\unit[9]{K}$ to $\unit[12.7]{K}$ but in the theoretical phase diagram, the temperature range where both phases appear goes from $\unit[0]{K}$ until near $\unit[17]{K}$.
Figures~\ref{f:phase_diagrams}a) and b) show the phase diagram of the system in 2D for distributed 
and localized moments, respectively. By performing simulations for the 3D system, we obtained the same Curie temperature for the localized moments as the 2D case but the phase transition to the ferromagnetic ordering was occurring at very small magnetic fields (around $\unit[15-20]{mT}$). In the 3D simulations for the distributed moments, the calculations indicate that the same phase transitions for skyrmionic and cycloidal regions occur at the same values for the temperature and external magnetic field but the transition to the paramagnetic ordering arises at higher Curie temperature. In Fig.~\ref{f:tubes} in the SI are also shown eight layers of the 3D simulation for the distributed moments. The figure emphasizes the tubular quasi-2D nature of skyrmions in GaV$_4$S$_8$.

\section{Conclusions}

In this theoretical study of the lacunar spinel GaV$_4$S$_8$, we find that the Dzyaloshinskii-Moriya interaction (DMI) calculated from first principles and the associated micromagnetic energy reflect the $C_{3\nu}$ crystal symmetry and support the formation of N\'eel skyrmions, previously reported for this system [\onlinecite{Kezsmarki2015}]. In contrast to previous works [\onlinecite{Zhang2017},\onlinecite{Nikolaev2019}], we obtain a detailed picture of the magnetic interactions, both between individual V sites and between different V$_4$ clusters, not just the nearest neighbors.

Electronic correlations are important in this multiferroic system, since, on the one hand, they open up a semiconducting band gap $\sim\!\!\unit[100]{meV}$ and, on the other hand, they enhance significantly the energy scale of the Dzyaloshinskii-Moriya interactions (spiralization $D$ in Eqn.~(\ref{e:spiralization_matrix})) relative to the Heisenberg exchange (spin stiffness $A$ in Eqn.~(\ref{e:spin_stiffness})). In particular, within the generalized-gradient approximation we find a smaller $A/D$ ratio of $\unit[7.4]{nm}$ for $U=\unit[2]{eV}$ (moderate correlations, localized-moment state) compared to the pure DFT result $A/D=\unit[26.3]{nm}$, even though the absolute value of $D$ is reduced by correlations. We believe that this behavior is related to the opening of the band gap and redistribution of the magnetic density in the V$_4$ cluster. Compared to the localized-moment state, the ratio $A/D=\unit[5.0]{nm}$ for the distributed-moment configuration is remarkably smaller, while the spin stiffness $A$ is a factor of 6 larger, allowing to distinguish these two cluster states based on the predicted magnetic properties.

Based on our micromagnetic simulations using our computed first-principles parameters, we conclude that a small $|A/D|$ ratio is important for stabilizing N\'eel skyrmions with a size $\unit[13-27]{nm}$ close to the measured one ($a_\mathrm{sky}=\unit[22.2]{nm}$, [\onlinecite{Kezsmarki2015}]), while the value of the spin stiffness $A$ determines the critical temperature for magnetic order. Although the uniaxial magnetic anisotropy is weak, it has a considerable effect on the magnetic phase diagram (Fig.~\ref{f:phase_diagrams}). With the literature values of the anisotropy, we find that the distributed-moment configuration, where all four V sites in each V$_4$ cluster have sizeable moments, describes much better the magnetic properties and textures in GaV$_4$S$_8$ for the experimentally studied temperature and magnetic field range. 

We note that there is a difference between the atomistic and micromagnetic results for this system. For example, in Fig.~\ref{f:M_vs_T_U=2} the Curie temperature, $T_C$, of the two different types of electronic (and magnetic) configurations are $\unit[15]{K}$ and $\unit[23]{K}$, while the respective estimates from Fig.~\ref{f:phase_diagrams} are $\unit[4]{K}$ and $\unit[15]{K}$, meaning a shift around $\unit[10]{K}$ between the 3D atomistic and 2D micromagnetic results. We have also made 3D micromagnetic simulations, which showed somewhat larger $T_C$ for the distributed-moment cluster, but the same $T_C$ for the localized-moment cluster. Overall, it is from these types of simulations difficult to pin-point an ordering temperature with an accuracy of a few kelvin,\cite{Eriksson2017} so that from these comparisons it is difficult to identify which electronic configuration (distributed or localized moments) is relevant for this system. However, magnetic textures at lower temperatures are more faithfully reproduced by the type of calculations/simulations presented here\cite{Eriksson2017}, and for this reason we argue that the four-site, distributed moment configuration in GaV$_4$S$_8$, which is crucial to reproduce the magnetic properties, represents the correct electronic configuration of this material.

It should be noted that the localized-moment configuration has a lower energy in the DFT calculations presented here, which seems to contradict the conclusion that the distributed moment configuration is the relevant one for GaV$_4$S$_8$. However, the energy difference between the types of configurations is not large and it is possible that dynamical correlations may change the balance so that the distributed-moment case becomes lower in energy than the local moment configuration. In view of the calculated magnetic phase diagrams (Fig.~\ref{f:phase_diagrams}) and their comparison with experiment, in particular, the observation of N\'eel skyrmions suggests strongly that the V$_4$ clusters are in the distributed-moment state. However, given the closeness of the different electronic configurations, and their distinctly different magnetic states, we speculate that the excitation spectra of GaV$_4$S$_8$ should be particularly interesting, both from traditional electron- and x-ray spectroscopic methods, as well as magnetic excitations, e.g., as provided by inelastic neutron scattering experiments. Experimental work is necessary to map out the complexities of the here proposed electronic and magnetic configurations of the V-based lacunar spinel GaV$_4$S$_8$.

\medskip
\textbf{Author contributions} \par
VB designed the study, performed all density functional theory calculations and a large part of micromagnetic simulations, wrote most of the manuscript and prepared figures 1--6 and 8--11. NS did further more extensive micromagnetic simulations to construct the detailed phase diagrams and prepared figures 7, 12 and 13 as well as wrote an accompanying text. MP helped NS with running some of the simulations and writing of the manuscript. AD contributed by discussing and analyzing the results and editing of the manuscript. OE helped with planning the theoretical work and analyzing the results, and contributed to the writing and editing of the manuscript. All authors have read and approved the final manuscript.

\medskip
\textbf{Competing interests} \par
All authors declare no financial or non-financial competing interests.

\medskip
\textbf{Acknowledgements} \par

\begin{figure*}
{\centering
\includegraphics[width=0.99\textwidth]{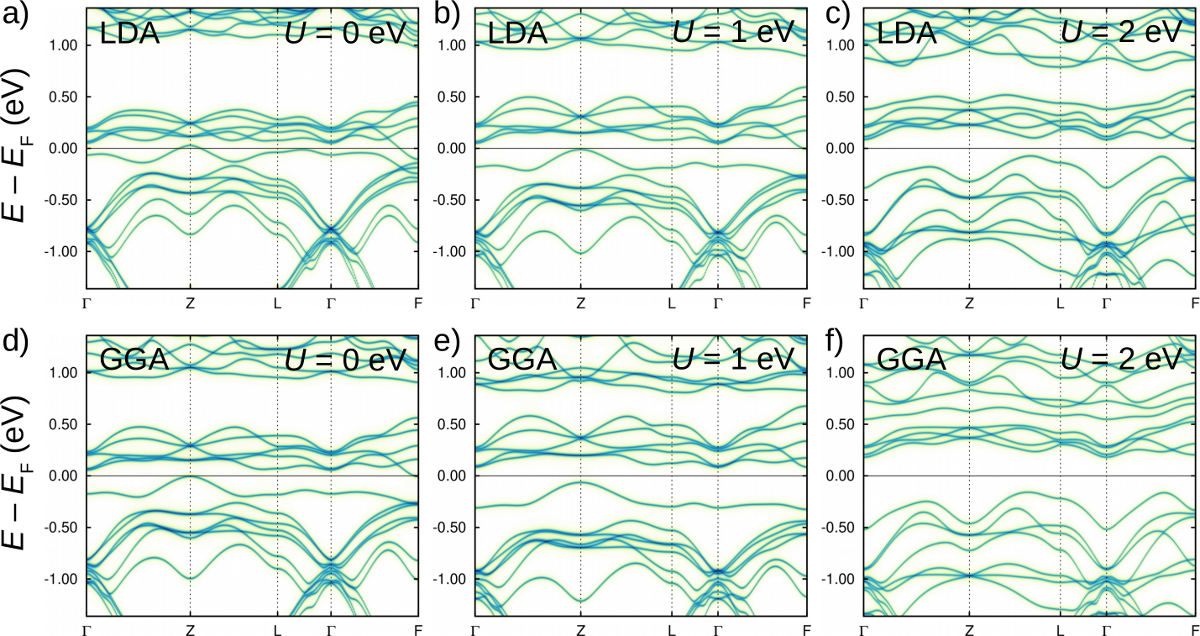}
}
\vspace{-10pt}
\caption{Electronic band structure of GaV$_4$S$_8$ calculated within density functional theory using LDA (a-c) and GGA (d-f) at different electronic correlation strengths characterized by the parameter $U$ in the DFT+$U$ scheme. For $U=\unit[0]{eV}$ the system is metallic but becomes semiconducting for $U>\unit[1]{eV}$ in LDA and for smaller $U>\unit[0]{eV}$ in GGA. For $U=\unit[2]{eV}$, the results for the distributed-moment state are shown; this state is higher in energy than the localized-moment state, but both states show similar band gaps.
}
\label{f:electronic_bands_GaV4S8}
\end{figure*}

This work was financially supported by the Knut and Alice Wallenberg Foundation through grant numbers 2018.0060, 2021.0246, and 2022.0108, and G\"oran Gustafsson Foundation (recipient of the ``small prize'': Vladislav Borisov). Olle Eriksson also acknowledges support by the Swedish Research Council (VR), the Foundation for Strategic Research (SSF), the Swedish Energy Agency (Energimyndigheten), the European Research Council (854843-FASTCORR), eSSENCE and STandUP. Anna Delin acknowledges financial support from Vetenskapsrådet (VR)(grant numbers VR 2016-05980 and VR 2019-05304). The computations/data handling were enabled by resources provided by the Swedish National Infrastructure for Computing (SNIC) at the National Supercomputing Centre (NSC, Tetralith cluster) partially funded by the Swedish Research Council through grant agreement no.\,2018-05973 and by the National Academic Infrastructure for Supercomputing in Sweden (NAISS) at the National Supercomputing Centre (NSC, Tetralith cluster) partially funded by the Swedish Research Council through grant agreement no.\,2022-06725. The funder played no role in study design, data collection, analysis and interpretation of data, or the writing of this manuscript. Structural sketches in Figs.~\ref{f:spinel_structure} and \ref{f:DM_interactions} have been produced by the \textsc{VESTA3} software \cite{vesta}. Figures~6 and 13 are produced using the \textsc{ParaView} software \cite{paraview}.

\appendix

\section{Band structures}

\begin{figure}
{\centering
\includegraphics[width=0.99\columnwidth]{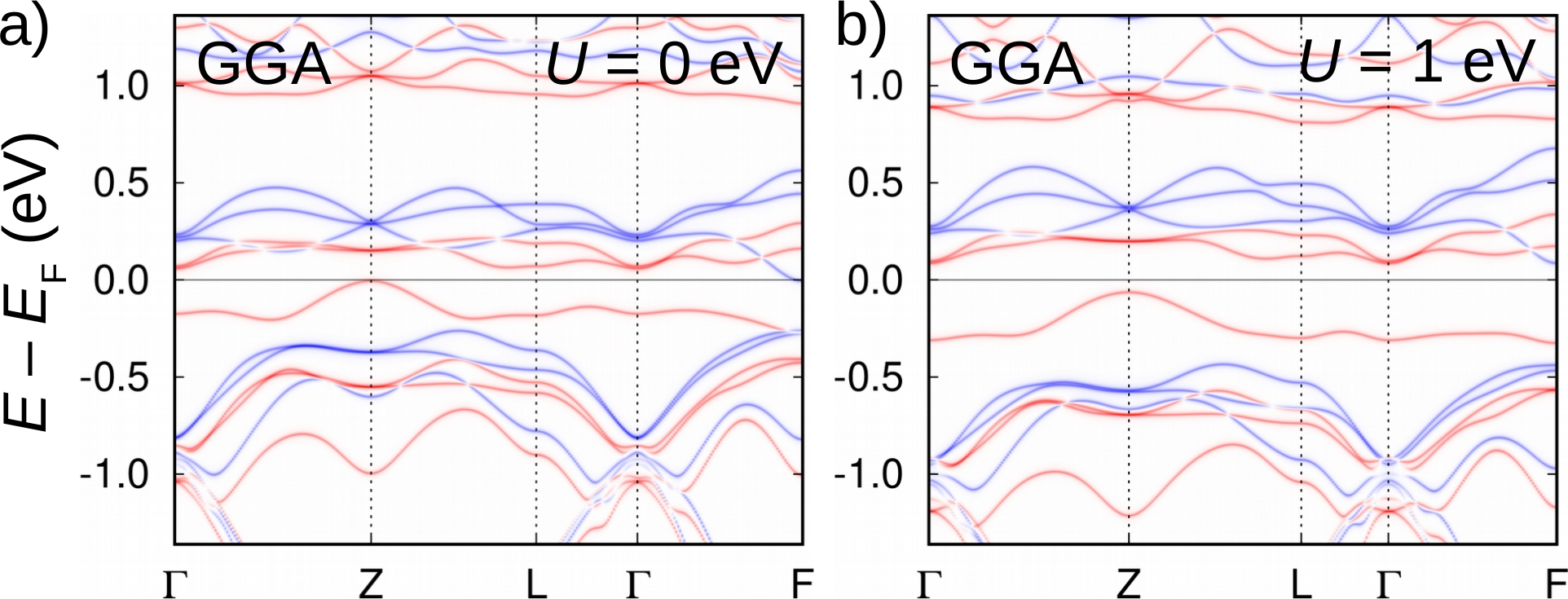}
}
\vspace{-10pt}
\caption{Spin-resolved electronic band structure of GaV$_4$S$_8$ calculated within density functional theory in the GGA and GGA+$U$ ($U=\unit[1]{eV}$) formalisms. Red and blue curves correspond to spin-up and down $d$ states projected on the V sites, and the V$_4$ clusters are in the localized-moment configuration (Fig.~\ref{f:spinel_structure}b in the main text).
}
\label{f:spin_polarized_band_structures}
\end{figure}

The comparison of electronic band structures of lacunar spinel GaV$_4$S$_8$ is shown in Fig.~\ref{f:electronic_bands_GaV4S8} where the calculation results obtained within the local-density (LDA) and generalized-gradient approximations (GGA) are collected. The effects of electronic correlations missing in the pure DFT are estimated by including DFT+$U$ corrections with varying $U$ parameter, which characterizes the strength of correlations. While the system is erroneously described as metallic by pure DFT ($U=\unit[0]{eV}$), DFT+$U$ corrections help to open the electronic band gap for $U>\unit[1]{eV}$ (LDA) and already at smaller $U$ values for GGA. The resulting gap is of the order of several hundred meV, in acceptable agreement with experiment. To give more details, Fig.~\ref{f:spin_polarized_band_structures} shows the spin-resolved bands for DFT and DFT+$U$ with $U=\unit[1]{eV}$. Overall, it appears that both the spin-up and spin-down bands are shifted mostly equally by the correlation corrections.

For $U<\unit[2]{eV}$, our calculations converge mostly to the localized-moment state of V$_4$ clusters (Fig.~\ref{f:spinel_structure}b in the main text), but for large $U$ values it is possible to stabilize also the distributed-moment state (Fig.~\ref{f:spinel_structure}c in the main text). The band structures for that configuration are shown in Fig.~\ref{f:electronic_bands_GaV4S8}c,f and are different from the localized-moment results, but the semiconducting band gap is similar for both V$_4$ cluster configurations. However, the absence of detailed ARPES data in the literature prevents the prediction of the V$_4$ cluster state based solely on the electronic properties.

\vspace{-15pt}
\section{Atomistic spin dynamics}

\begin{figure}
\vspace{-10pt}
{\centering
\includegraphics[width=0.9\columnwidth]{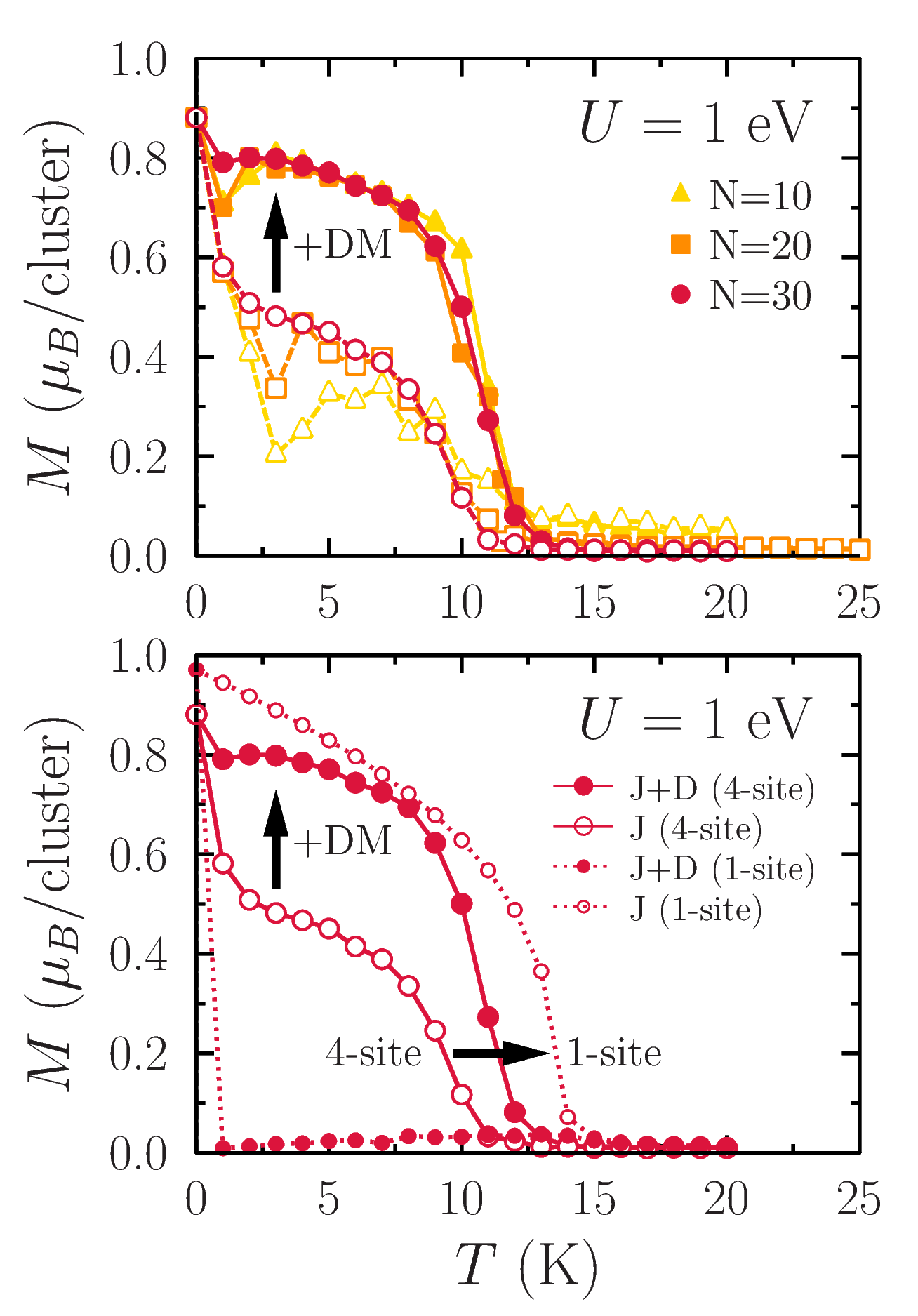}
}
\vspace{-10pt}
\caption{a) Temperature dependence of the magnetization $M(T)$ obtained from classical atomistic Monte Carlo simulations based on first-principles Heisenberg $J_{ij}$ and DM $\vec{D}_{ij}$ intersite interactions (filled symbols) and Heisenberg interactions only (empty symbols), obtained within GGA+$U=\unit[1]{eV}$. Different supercell sizes $N^3$ are used in the simulations. b) $M(T)$ for $N=30$ obtained using the 4-site (solid lines) and 1-site (dashed lines) models, where in the latter case the intracluster degrees of freedom are not resolved.}
\vspace{-5pt}
\label{f:M_vs_T}
\end{figure}
When performing atomistic spin dynamics simulations of the lacunar spinel using the calculated magnetic interactions, we verified that the size effects do not change the results considerably. In Fig.~\ref{f:M_vs_T}a, the temperature-dependent magnetization is plotted for simulation $(N\times N\times N)$ cells of increasing size $N=10, 20, 30$. The variations due to the cell size are apparently very small when all the interactions are included. If only Heisenberg interaction is taken into account (open symbols in Fig.~\ref{f:M_vs_T}a), then only the $30\times 30\times 30$ or larger cells provide sufficiently converged results.

Interestingly, the Dzyaloshinskii-Moriya interaction appears to change dramatically the temperature-dependence of the magnetization. This can be seen from the Monte-Carlo results obtained using only the Heisenberg interactions $J_{ij}$ (open symbols in Fig.~\ref{f:M_vs_T}a) compared to the simulations which include also the DM interaction $\vec{D}_{ij}$ (filled symbols in Fig.~\ref{f:M_vs_T}a). The results suggest that the DMI enhances the long-range ferromagnetic order in a temperature range between 0 and $\unit[12]{K}$, which is rather surprising considering that DMI usually tries to make the magnetic structure more non-collinear. In this respect, GaV$_4$S$_8$ spinel contrasts many other known magnets and this aspect deserves further analysis which will be done in the future work.

Another important point is the difference between atomistic simulations based on inter-site (4-site model) or inter-cluster (1-site model) interactions, i.e. whether the individual sites or whole V$_4$ clusters are considered as the elementary magnetic units. From Fig.~\ref{f:M_vs_T}b, we can clearly see that neglecting the intracluster degrees of freedom by considering a 1-site spin model with effective interactions (eqn.~(\ref{e:effective_interactions})) changes the behavior of the magnetization $M(T)$ significantly, as obtained in atomistic spin dynamics simulations, even when the DM interaction is not included. On the other hand, if the 1-site model includes the effective DM interaction, defined similarly to eqn.~(\ref{e:effective_interactions}), we find that the long-range magnetic order is basically totally suppressed at any temperature above $\unit[1]{K}$. In micromagnetic simulations, we do not see this drawback of the 1-site model. These findings suggest that the modelling of intracluster degrees of freedom in lacunar spinel GaV$_4$S$_8$ is a subtle issue and has to be considered in greater detail in future studies.

\section{Micromagnetic simulations}

\begin{figure}
\vspace{15pt}
{\centering
\includegraphics[width=0.99\columnwidth]{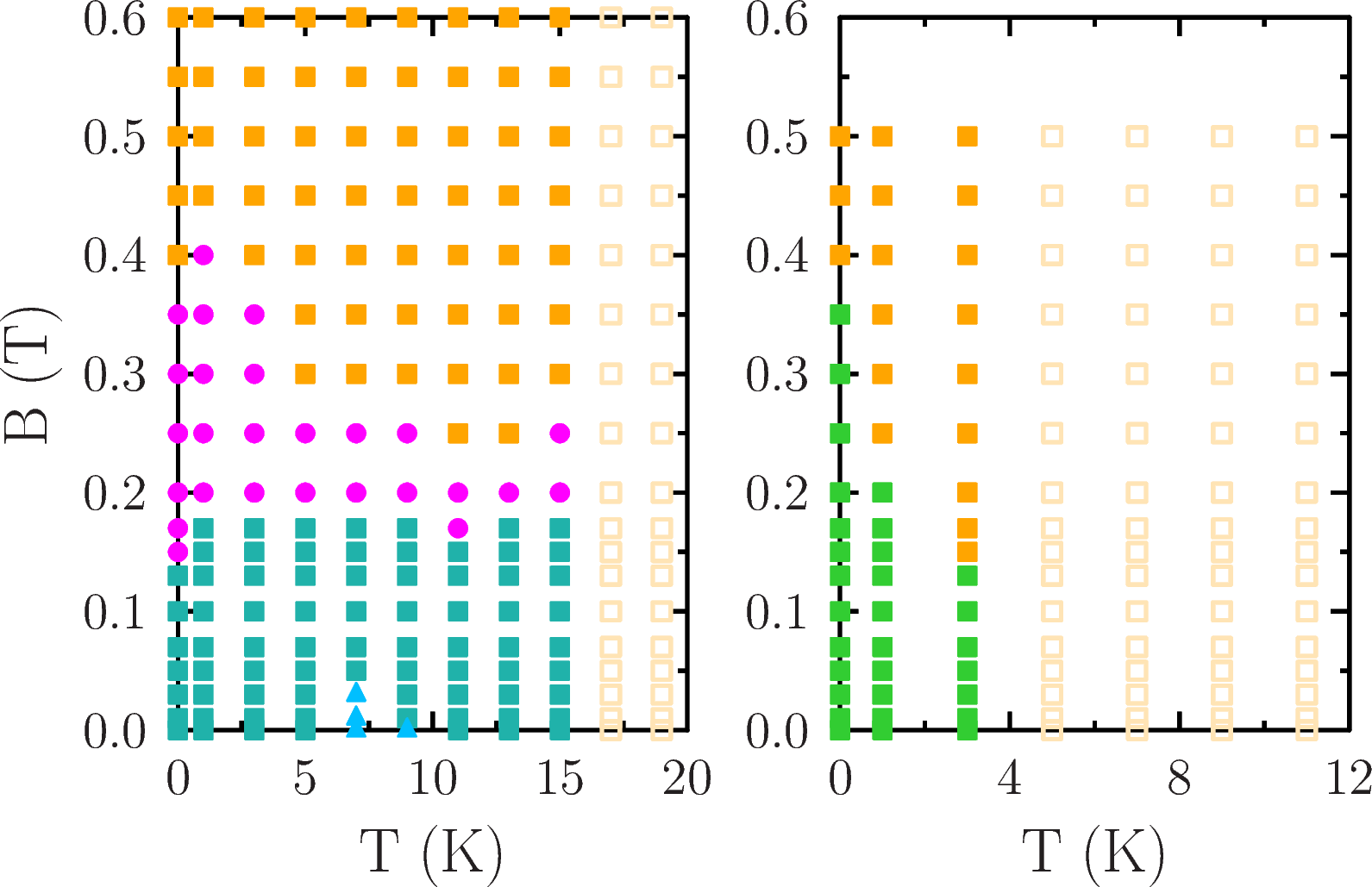}
}
\vspace{-15pt}
\caption{Original theoretical data points used for constructing the phase diagrams of GaV$_4$S$_8$ in Fig.~\ref{f:phase_diagrams} of the main text (the color code is the same).}
\label{f:phase_diagrams_data}
\end{figure}
\begin{figure*}[h]
{\centering
\includegraphics[width=0.99\textwidth]{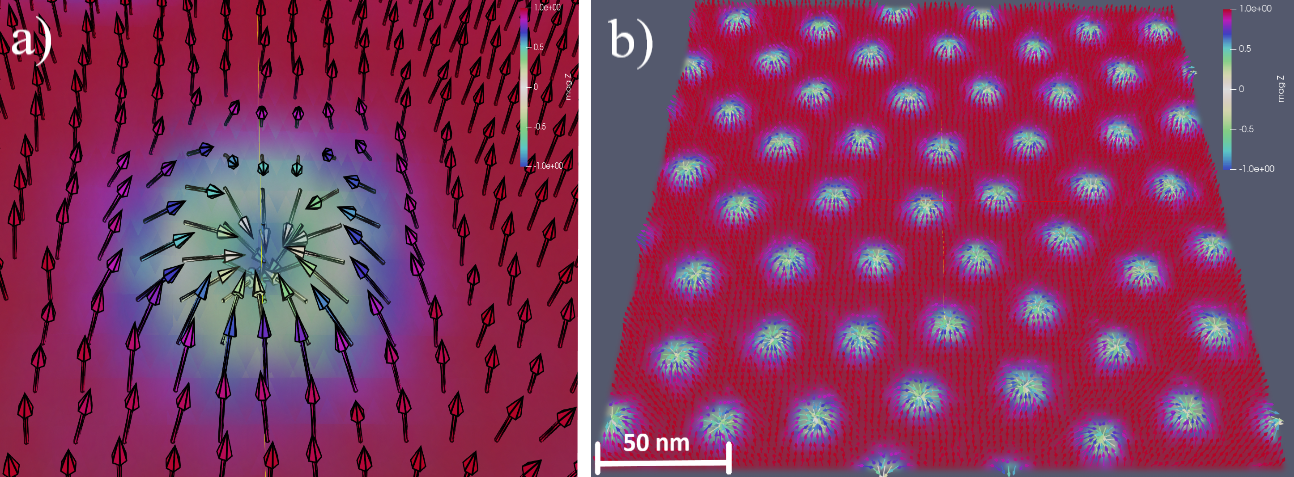}
}
\vspace{-10pt}
\caption{Magnetic texture representing a) N\'eel-type skyrmion and b) a skyrmion lattice. The color bar represents the $z$ component of the magnetization. The figure has been plotted by using the $\mu$ASD software\cite{multiscale}.}
\label{f:neel_skyrmion}
\end{figure*}

Based on the micromagnetic simulations for different temperatures and magnetic field strengths, we constructed the phase diagrams for GaV$_4$S$_8$ with two different V$_4$ cluster states, shown in Fig.~\ref{f:phase_diagrams} in the main text. The original data points that we obtained in those \textsc{mumax3} simulations are depicted here in Fig.~\ref{f:phase_diagrams_data} using the same color code for convenience. It is worth noting that in our \textsc{UppASD} simulations using the multiscale module $\mu$\textsc{ASD}\cite{multiscale} we observed skyrmion lattices in some conditions (example shown in Fig.~\ref{f:neel_skyrmion}).

\begin{figure*}[h]
{\centering
\includegraphics[width=0.99\textwidth]{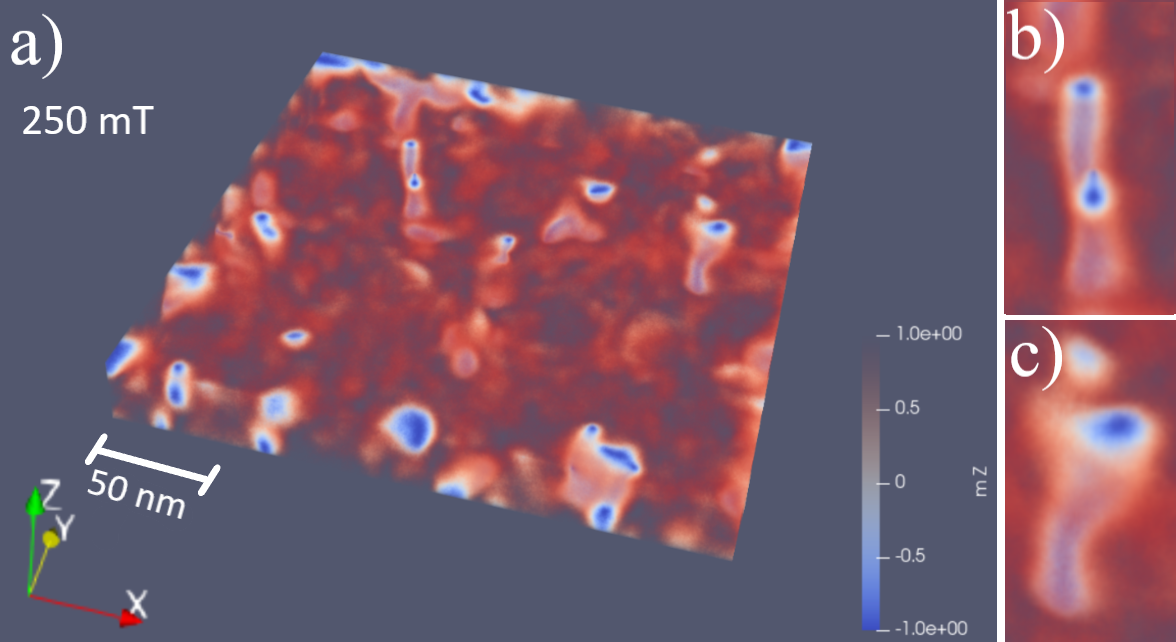}
}
\vspace{-10pt}
\caption{Skyrmion tubes for a) a 3D simulation for distributed-moment V$_4$ cluster state and simulation cell with eight layers along the $z$-direction and b), c) zoom-ins of skyrmion tubes. The color bar represents the $z$ component of the magnetization.}
\label{f:tubes}
\end{figure*}

For selected points in \textsc{mumax3} phase diagrams, we have verified the effect of cell size along the $z$-direction, i.e.~we compared the 2D and 3D simulation results. One example for external field of $\unit[250]{mT}$ is shown in Fig.~\ref{f:tubes} based on the simulation with 8 layers along the $z$-direction. It turns out that the phase diagrams do not change qualitatively, but the Curie temperature for the distributed-moment configuration increases up to $\unit[25]{K}$, resulting in a shift of the boundary to the paramagnetic phase in Fig.~\ref{f:phase_diagrams_data}a and agreeing better with the atomistic simulations in Fig.~\ref{f:M_vs_T_U=2} in the main text. For the localized-moment state (Fig.~\ref{f:phase_diagrams_data}b), however, there is no visible shift of the Curie temperature in the 3D simulations.

\medskip
\bibliographystyle{prb-titles}
\bibliography{main}

\end{document}